\documentclass[12pt]{article}
\usepackage[spanish,es-tabla]{babel}
\usepackage{natbib}
\usepackage{comment}
\usepackage{float}
\usepackage[hidelinks]{hyperref}
\usepackage{mathrsfs}
\usepackage{enumitem}
\usepackage[font={small,it}]{caption}
\usepackage{amsmath,amsfonts,amsthm,amssymb}
\usepackage{bm,rotating,multirow,dsfont,graphicx}
\usepackage[usenames, dvipsnames]{color}
\usepackage{url}
\usepackage{multicol}
\usepackage{multirow}
\usepackage[T1]{fontenc}
\usepackage{flafter}
\usepackage{appendix}
\usepackage{subfigure}
\usepackage{xcolor}
\usepackage{ragged2e}
\usepackage{listings}

\makeatletter
\def\hlinewd#1{%
	\noalign{\ifnum0=`}\fi\hrule \@height #1 %
	\futurelet\reserved@a\@xhline}
\makeatother
\usepackage[all,cmtip]{xy}

\def\frac#1#2{{{#1}\over{#2}}}

\def\@roman#1{\romannumeral #1}

\begin{document}

\renewcommand{\refname}{Referencias}

\def\spacingset#1{\renewcommand{\baselinestretch}{#1}\small\normalsize}\spacingset{1}

\title{Modelamiento Bayesiano de las preferencias políticas \\ del Senado de Colombia 2006-2010: \\ conducta electoral y parapolítica}

\date{}
	
\author{
    Juan Valero\footnote{Universidad Nacional de Colombia. Email: juavalerosi@unal.edu.co .},
    Juan Sosa\footnote{Universidad Nacional de Colombia. Email: jcsosam@unal.edu.co .}, 
    Carolina Luque\footnote{Universidad Ean. Email: cluque2.d@universidadean.edu.co .}
}	 

\maketitle

\begin{abstract} 
En este trabajo se aplica por primera un modelo de votación espacial Bayesiano para caracterizar el comportamiento legislativo del Senado de la República de Colombia del periodo 2006-2010. El análisis se realiza con base en las votaciones nominales plenarias del Senado. La estimación del modelo se hace mediante algoritmos de cadenas de Markov Monte Carlo. Los puntos ideales estimados proveen evidencia empírica que sustenta un rasgo latente no ideológico (oposición--no oposición) subyacente a la votación de los senadores. Adicionalmente, se analiza la relación entre el escándalo de la parapolítica y el comportamiento legislativo de los senadores mediante un modelo logístico tanto Bayesiano como frecuentista. Los resultados indican una relación significativa entre estar o haber estado involucrado con el escándalo de la parapolítica y el comportamiento legislativo de los senadores del periodo 2006-2010.
\end{abstract}

\noindent
{\it Palabras clave:} Estimador de punto ideal Bayesiano; Métodos de cadenas de Markov de Monte Carlo; Parlamentos desequilibrados; Parapolítica; Votos nominales.

\spacingset{1.1} 


\newpage

\section{Introducción}

Los modelos de votación espacial están basados en la creencia popular de que un sujeto toma decisiones políticas basado en su ideología \citep{curini2020sage}. Los \textit{puntos ideales} estimados por el modelo de votación espacial suelen ser interpretados como la forma en que se captura la ideología política del sujeto \citep{yu2020spherical}. La estimación del punto ideal se hace a partir de las decisiones que el actor toma en un conjunto de propuestas de votación. Estos modelos se usan para analizar el comportamiento electoral en diferentes cuerpos legislativos como lo son el Senado, las Cámaras, los Consejos, las Cortes, etc \citep{clinton2004statistical}.

Entre los modelos de votación espacial destaca principalmente el de la estimación del punto ideal Bayesiano. La estimación del punto ideal Bayesiano para los modelos de votación espacial fue inicialmente promulgado por Clinton y colaboradores en 2004 \citep{curini2020sage}. Este modelo es lo suficientemente flexible como para ser usado en cualquier cuerpo legislativo \citep{clinton2004statistical}.

En Colombia este modelo fue implementado para el Senado de la República 2010-2014 en \cite{luque2022bayesian}. En este trabajo se lleva a cabo el proceso de consolidación de la base de datos que contiene las votaciones legislativas y se hacen recomendaciones para caracterizar los patrones en las preferencias políticas de los senadores para otros periodos, como por ejemplo el de 2006-2010.

En esencia, en el presente documento se emula lo hecho en \cite{luque2022bayesian} para el Senado de la República 2006-2010. Por tal motivo, los objetivos primordiales son i. consolidar la base de datos de senadores y votaciones de plenaria del periodo 2006-2010, ii. estimar los puntos ideales de los senadores a partir de los registros de votación plenaria, iii. encontrar patrones en las preferencias políticas que subyacen a la votación, y iv. determinar si el escándalo de la parapolítica tuvo efecto en el comportamiento legislativo de los senadores.

Así, este documento se estructura como sigue: La Sección 2 describe el proceso de conformación de la base de datos. La Sección 3 presenta un análisis descriptivo, donde buena parte del análisis está enfocado en caracterizar el efecto del escándalo de la parapolítica en la asistencia, la participación, y la abstención de los senadores. La Sesión 4 recopila los detalles acerca de la implementación del modelo. La Sesión 5 analiza las estimaciones de los puntos ideales, donde se dedica una subsección exclusivamente a analizar la relación entre el escándalo de la parapolítica y el comportamiento legislativo de los senadores. Finalmente, la Sesión 6 discute los resultados del trabajo y algunas futuras alternativas de investigación.

\section{Estimación Bayesiana del Punto Ideal}

En el modelo de votación espacial los datos consisten de $n$ actores que votan en $m$ listas de votación diferentes. Cada voto es una elección entre dos opciones: ``SI'' o ``NO". El voto del $i$-ésimo actor en la lista $j$-ésima se representa por $y_{i,j}$ $i=1,\cdots,n$, $j=1,\cdots,m$, donde esta variable toma el valor $1$ si el voto es favor y $0$ si el voto es en contra. Las alternativas de votación se representan como puntos en el \textit{espacio político} ($\mathbb{R}^d$). Para la $j$-ésima votación se usa $\mathbf{p}_j$ para el punto que representa el ``SI'' y $\mathbf{q}_j$ para el punto que representa el ``NO". La preferencia de cada actor entre estas dos opciones está determinada por la cercanía de estos puntos a la preferencia política del legislador, la cual está dada por un factor no observado $\boldsymbol{\beta}_i$ conocido como \textit{punto ideal}.
De otra parte, hay opciones no euclidianas para el espacio político, por ejemplo, en \cite{yu2020spherical} se proponen variedades diferenciales esféricas como alternativa cuando el mecanismo subyacente de generación de los datos implica fenómenos en extremo no lineales.

En el modelo de votación espacial se considera que es más probable que los actores elijan la opción más cercana a su ideal político (punto ideal) que la opción que se encuentra más alejada de su ideal político \citep{carroll2013structure}. Lo anterior se modela mediante el uso de funciones de utilidad estocásticas, las cuales dependen tanto de la cercanía del punto ideal a la opción de votación como de un error aleatorio \citep{poole2005spatial}.
Se asume que la manera en que los actores votan está basada en su punto ideal de acuerdo con funciones de utilidad dadas por:
$$U_i(\mathbf{p}_j)=-\|\mathbf{p}_j-\boldsymbol{\beta}_i\|^2+\eta_{i,j} \qquad \text{y} \qquad U_i(\mathbf{q}_j)=-\|\mathbf{q}_j-\boldsymbol{\beta}_i\|^2+\nu_{i,j}\,.$$
Las utilidades del actor i-ésimo al votar a favor o en contra están dadas por $U_i(\mathbf{p}_j)$ y $U_i(\mathbf{q}_j)$, respectivamente. Se asume que $\eta_{i,j}$ y $\nu_{i,j}$ son variables aleatorias independientes tal que el valor esperado y la varianza de $\eta_{i,j}-\nu_{i,j}$ son $0$ y $\sigma_j^2$, respectivamente.

La forma de la función de utilidad tiene implícitos algunos supuestos:  lo único relevante en el espacio político es la distancia entre los puntos ideales y las opciones de votación, la utilidad de un voto a favor (en contra) decrece a medida que $\|\mathbf{p}_j-\boldsymbol{\beta}_i\|$ ($\|\mathbf{q}_j-\boldsymbol{\beta}_i\|$) aumenta, y las preferencias son simétricas \cite{curini2020sage}.

El actor $i$-ésimo vota a favor de la lista $j$-ésima si y sólo si $U_i(\mathbf{p}_j)>U_i(\mathbf{q}_j)$. De lo anterior se tiene que: 
\begin{equation*}
y_{i,j}\mid \mathbf{p}_j,\mathbf{q}_j,\sigma_j,\boldsymbol{\beta}_i = 
    \begin{cases}
        1, & \text{si } U_i(\mathbf{p}_j)-U_i(\mathbf{q}_j)>0 \\
        0, & \text{en caso contrario. } 
    \end{cases}
\end{equation*}
Por lo tanto, 
\begin{align*}
\textsf{Pr}(y_{i,j}=1\mid \mathbf{p}_j,\mathbf{q}_j,\sigma_j,\boldsymbol{\beta}_i) &=\textsf{Pr}(U_i(\mathbf{p}_j)-U_i(\mathbf{q}_j)>0)    
\end{align*}
Resolviendo la diferencia $U_i(\mathbf{p}_j)-U_i(\mathbf{q}_j)$, se obtiene que:
\begin{align*}
\textsf{Pr}(y_{i,j}=1\mid\mathbf{p}_j,\mathbf{q}_j,\sigma_j,\boldsymbol{\beta}_i)
= \textsf{Pr}(\epsilon_{i,j}<\mu_j+\boldsymbol{\alpha}^\textsf{T}_j\boldsymbol{\beta}_i) =\textsf{G}(\mu_j+\boldsymbol{\alpha}_j^\textsf{T}\boldsymbol{\beta}_i)\,,
\end{align*}
donde $\epsilon_{i,j}=(\nu_{i,j}-\eta_{i,j})/\sigma_j$, $\mu_j=(\mathbf{q}_j^\textsf{T}\mathbf{q}_j-\mathbf{p}_j^\textsf{T}\mathbf{p}_j)/\sigma_j$ y $\boldsymbol{\alpha}_j=2(\mathbf{p}_j-\mathbf{q}_j)/\sigma_j$. El parámetro $\mu_j$ representa el parámetro de aprobación, $\boldsymbol{\alpha}_j$ es el parámetro de discriminación y $\textsf{G}$ es una función de enlace apropiada. Para los detalles matemáticos \citep{luque2022bayesian}.

La distribución de la variable aleatoria $\epsilon_{i,j}$ determina el tipo de modelo. Por ejemplo, si $\epsilon_{i,j}$ tiene distribución normal estándar entonces \textsf{G} conduce a un modelo probit, en cuyo caso se tiene que
$$y^*_{i,j}=\mu_j+\boldsymbol{\alpha}_j^\textsf{T}\boldsymbol{\beta}_i+\epsilon_{i,j} \qquad \text{con} \qquad \epsilon_{i,j} \overset{\mathrm{iid}}{\sim} \textsf{N}(0,1)\,, $$
donde $y^*_{i,j}$ es el predictor lineal que impulsa la probabilidad de voto a favor y por lo tanto se tiene que: 
$$y_{i,j}\mid\mu_j, \boldsymbol{\alpha}_j,\boldsymbol{\beta}_i \overset{\mathrm{iid}}{\sim} \textsf{Bernoulli}(\textsf{G}(\mu_j+\boldsymbol{\alpha}_j^\textsf{T}\boldsymbol{\beta}_i)).$$
A diferencia de los modelos de regresión común y corrientes, tanto $\boldsymbol{\alpha}_j$ como $\boldsymbol{\beta}_i$ son desconocidos. Los vectores $\boldsymbol{\beta}_i$ están indexados con $i$ debido a que son parámetros asociados a los actores, mientras que los vectores $\boldsymbol{\alpha}_j$ están indexados con $j$ debido a que son parámetros asociados a las listas de votación. Lo anterior implica que hay $m$ parámetros $\boldsymbol{\alpha}$ y $n$ parámetros $\boldsymbol{\beta}$ por estimar.

\begin{figure}[H]
    \centering
    \includegraphics[scale=0.6]{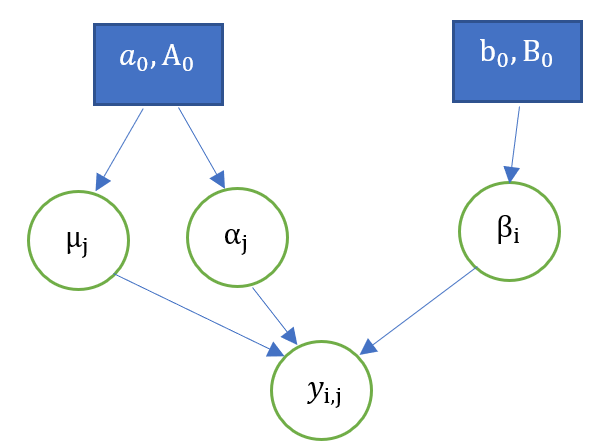}
    \caption{Grafo acíclico dirigido para el modelo de votación espacial.}
    \label{DAG}
\end{figure}

Sea $\mathbf{Y}$ la matriz de tamaño $n\times m$ de votos observados cuyo elemento $(i,j)$-ésimo es $y_{i,j}$, $\boldsymbol{\mu}=(\mu_1,\ldots,\mu_m)$, $\mathbf{A}$ la matriz $m\times d$ cuya $j$-ésima fila es $\boldsymbol{\alpha}^\textsf{T}_j$ y $\mathbf{B}$ la matriz de tamaño $n\times d$ cuya $i$-ésima fila es $\boldsymbol{\beta}^\textsf{T}_i$. Bajo la hipótesis de que los $y_{i,j}$ son intercambiables\footnote{El lector interesado en el concepto de intercambiabilidad puede encontrarlo en \cite{bernardo2009bayesian}.} dados $\mu_j$, $\boldsymbol{\alpha}_j$ y $\boldsymbol{\beta}_i$, se tiene que: 
$$\textsf{p}(\mathbf{Y}\mid\boldsymbol{\alpha},\mathbf{A},\mathbf{B})=\prod_{i=1}^n\prod_{j=1}^m \textsf{G}(\mu_j+\boldsymbol{\alpha}_j^\textsf{T}\boldsymbol{\beta}_i)^{y_{i,j}}[1-\textsf{G}(\mu_j+\boldsymbol{\alpha}_j^\textsf{T}\boldsymbol{\beta}_i)]^{1-y_{i,j}}\,,$$
donde $\mathbf{Y}$ es la matriz de tamaño $n\times m$ de votos observados cuyo elemento $(i,j)$-ésimo es $y_{i,j}$, $\boldsymbol{\mu}=(\mu_1,\ldots,\mu_m)$, $\mathbf{A}$ es la matriz $m\times d$ cuya $j$-ésima fila es $\boldsymbol{\alpha}^\textsf{T}_j$ y $\mathbf{B}$ es la matriz de tamaño $n\times d$ cuya $i$-ésima fila es $\boldsymbol{\beta}^\textsf{T}_i$. En la Figura \ref{DAG} se muestra el grafo acíclico dirigido que exhibe la estructura del modelo de votación espacial. 
Para completar la especificación del modelo, se usan distribuciones previas normales para los parámetros $\mu_j$, $\boldsymbol{\alpha}_j$ y $\boldsymbol{\beta}_i$. Así, 
$$(\mu_j,\boldsymbol{\alpha}_j)\mid \boldsymbol{\alpha}_0,\mathbf{A}_0 \overset{\mathrm{iid}}{\sim} \textsf{N}(\boldsymbol{\alpha}_0,\mathbf{A}_0) \qquad \text{y} \qquad  \boldsymbol{\beta}_i\mid \mathbf{b}_i,\mathbf{B}_i \overset{\mathrm{iid}}{\sim} \textsf{N}(\mathbf{b}_i,\mathbf{B}_i)$$
donde $\boldsymbol{\alpha}_0,\mathbf{A}_0,\mathbf{b}_i,\mathbf{B}_i$ son los hiperparámetros del modelo. Los vectores $\boldsymbol{\alpha}_0,\mathbf{b}_i$ son las medias y $\mathbf{A}_0, \mathbf{B}_i$ son las matrices de varianzas y covarianzas. Los valores que se usan usualmente para los hiperparámetros son: $\boldsymbol{\alpha}_0=\mathbf{0}$, $\mathbf{A}_0=25\mathbf{I}$, $\mathbf{b}_0=\mathbf{0}$ y $\mathbf{B}_i=\mathbf{I}$ \citep{luque2022bayesian,clinton2004statistical}.

El objetivo primordial es estimar los puntos ideales $\boldsymbol{\beta}_1,\ldots,\boldsymbol{\beta}_n$ con el fin de encontrar y describir los rasgos latentes que subyacen a la votación. Los detalles del proceso de estimación de los parámetros se encuentra en \cite{luque2022bayesian}, donde se muestran procesos basados en cadenas de Markov para explorar la distribución posterior dada por $\textsf{p}(\mathbf{Y}\mid\mathbf{\alpha},\mathbf{A},\boldsymbol{B}) \propto \textsf{p}(\mathbf{Y}\mid\boldsymbol{\alpha},\mathbf{A},\mathbf{B})\,\textsf{p}(\boldsymbol{\alpha},\mathbf{A},\mathbf{B})$.

Es importante resaltar que tal y como está el modelo, los parámetros beta no son identificables debido a que si se tiene una matriz ortogonal $\mathbf{Q}$ (i.e., $\mathbf{Q}^{\textsf{T}}=\mathbf{Q}^{-1}$) y se aplica la transformación lineal inducida por esta matriz al espacio político, entonces las distancias entre alternativas de votación y puntos ideales permanecen constantes, y en tal caso los datos no pueden distinguir entre los diferentes valores de los parámetros.

\section{Conformación de la base de datos}

El registro de las votaciones de los proyectos de ley del Congreso de la República de Colombia se obtiene de Congreso Visible (\url{https://congresovisible.uniandes.edu.co/}).
El caso que se trabaja aquí es el del Senado de la República 2006-2010 donde los actores son los senadores y las listas son las votaciones en plenaria durante le cuatrenio 2006-2010.

Las abstenciones, fallas, o no inclusión en la lista de votación, son datos faltantes. Siguiendo a \cite{luque2022bayesian}, los datos faltantes se puede remover sin afectar los resultados. Estos autores realizan un análisis de sensitividad a diferentes tasas de datos faltantes para un parlamento desequilibrado, tal como el de este caso, demostrando que eliminar tales datos no tiene un impacto sustancial sobre los resultados.

Para determinar los senadores del periodo 2006-2010 se usó como fuentes principales Congreso Visible junto con Wikipedia. Aunque Wikipedia es una enciclopedia libre editada de manera colaborativa, se destaca por ser una de las mayores fuentes de información en la red que con el paso del tiempo ha implementado varias medidas con el fin de brindar mayor fiabilidad en sus contenidos\footnote{ver \url{https://www.elheraldo.co/tecnologia/que-tan-confiable-es-wikipedia-154671}}. Adicionalmente, también se utilizaron los principales medios de comunicación impresos y digitales colombianos tales como: Revista Semana, Periódico El Tiempo, El Espectador, entre otros, los cuales sirvieron para constatar la lista de senadores, el partido político al que pertenecían, los cambios de partido, las renuncias, la vinculación al escándalo de la parapolítica, etc.

Para el Senado del periodo de 2006-2010 fueron elegidos en las elecciones legislativas del año 2006 un total de 102 senadores de los cuales más de 30 se retiraron, fueron capturados (en su mayoría por parapolítica) o fallecieron. Razón por la cual hubo una gran cantidad de reemplazos durante ese cuatrenio. Se contabilizaron un total de 147 senadores en el transcurso de este periodo legislativo.
Luego de conformar la lista de senadores se hizo el cruce de esta con la lista de votaciones de los proyectos de ley del congreso con lo cual se obtuvo un total de 136 listas de votación en plenaria para el periodo en cuestión.

\section{Análisis descriptivo}

El Senado del periodo 2006-2010 corresponde al segundo mandato presidencial de Alvaro Uribe Vélez. El expresidente Uribe tuvo una alta aprobación por la opinión pública, lo cual también se reflejó en los apoyos con los que contó en el Congreso durante este periodo. La mayoría de los partidos eran afines al expresidente Uribe y su política de seguridad democrática. Durante este periodo hubo oposición por parte del partido Polo Democrático Alternativo y el Partido Liberal Colombiano, mientras que el partido MIRA fungió como independiente.

Durante este periodo fue que se destapó el escándalo de la parapolítica, lo cual produjo muchas renuncias y destituciones en el Congreso de la República. Además, para finales del año 2009 se aprobó una reforma política que permitía que los senadores cambiaran de partido político. Esto último fue usado por algunos senadores, que eran parte de partidos políticos muy afectados por el escándalo de la parapolítica, para moverse hacía otros partidos con el fin de huir del estigma de la parapolítica.

\subsection{Senadores y partidos políticos}

Los partidos que tuvieron senadores en el Congreso del periodo 2006-2010 fueron los siguientes: Partido de la U (PU), Conservador Colombiano (PC), Cambio Radical (CR), Colombia Viva (CV), Convergencia Ciudadana (CC), Alas Equipo Colombia (AEC), Colombia Democrática (CD), Polo Democrático Alternativo (PDA), Liberal Colombiano (PL), Movimiento MIRA (MIRA), Autoridades Indígenas de Colombia (AICO), y Alianza Social Indígena (ASI). De estos partidos, los 7 primeros eran parte de la coalición de partidos que apoyaban al entonces presidente Álvaro Uribe, el PDA y el PL conformaban la oposición, el MIRA era independiente, y los demás partidos hacían parte de las minorias.

\begin{table}[h]
\centering
\begin{tabular}{|c|c|c|c|c|c|c|c|}
\hline
Movimiento político & Número de curules & Número de senadores  \\ \hline
Alas Equipo Colombia & 5 & 8  \\ \hline
Movimiento Mira & 2 & 2  \\ \hline
Partido Conservador Colombiano & 18 & 24  \\ \hline
Convergencia Ciudadana & 7 & 11  \\ \hline
Partido de la U & 20 & 29  \\ \hline
Movimiento Colombia Viva & 2 & 6  \\ \hline
Polo Democrático Alternativo & 10 & 12  \\ \hline
Colombia Democrática & 3 & 8  \\ \hline
Partido Liberal Colombiano & 18 & 22  \\ \hline
Cambio Radical & 15 & 23  \\ \hline
Alianza Social Indígena & 1 & 1  \\ \hline
Autoridades Indígenas de Colombia & 1 & 1  \\ \hline
\end{tabular}
\caption{Número de curules por partido y número de senadores por partido durante el periodo 2006-2010.}
\label{tab:table0}
\end{table}

En el 2006 fueron elegidos 102 senadores de los cuales el $68\%$ eran miembros de los partidos de la coalición uribista, $27\%$ eran miembros del partido de oposición y los restantes eran miembros de partidos independientes o minorías. Durante los 4 años de este Senado se presentaron varias renuncias y destituciones, que en su mayoría estuvieron relacionadas con la parapolítica.

En la Tabla \ref{tab:table0} se encuentra el número de curules por movimiento político obtenidas para el periodo 2006-2010 y a su vez el número de senadores por movimiento político que fueron parte del Senado de la República durante el periodo 2006-2010. Los partidos con mayor número de curules fueron el PU, el PC, el PL y CR. Los partidos con menor número de curules fueron los partidos de minorías, el MIRA y CV. Los partidos CD y CV son los que presentan mayor diferencia entre su número de curules y de senadores, esto debido a que algunos de los senadores que entraron como reemplazo tuvieron que ser reemplazados. Es decir, hubo curules en estos partidos que llegaron a ser ocupadas por tres personas (en diferentes tiempos) durante el periodo en cuestión. Por otra parte, los únicos partidos que no tuvieron cambios de senadores fueron los partidos de minorías y el MIRA. El PDA solo tuvo un reemplazo, quien fue la senadora Gloria Isabel Cuartas Montoya, la cual reemplazó al Senador Jesús Antonio Bernal Amorocho a mediados de Mayo del 2010. El senador Guillermo Alfonso Jaramillo del PDA obtuvo su lugar en el senado mediante una demanda ante el Consejo de Estado, el senador Jaramillo había perdido en las elecciones del 2006 por 25 votos. En el año 2009 el Consejo de Estado le dio la razón al senador Jaramillo y de esta forma llegó al senado. No fue posible determinar si en este proceso algún senador fue destituido de su cargo para que así el senador Jaramillo pudiese tomar posesión de su curul.

\begin{figure}[!h]
    \centering
    \includegraphics[scale=0.62]{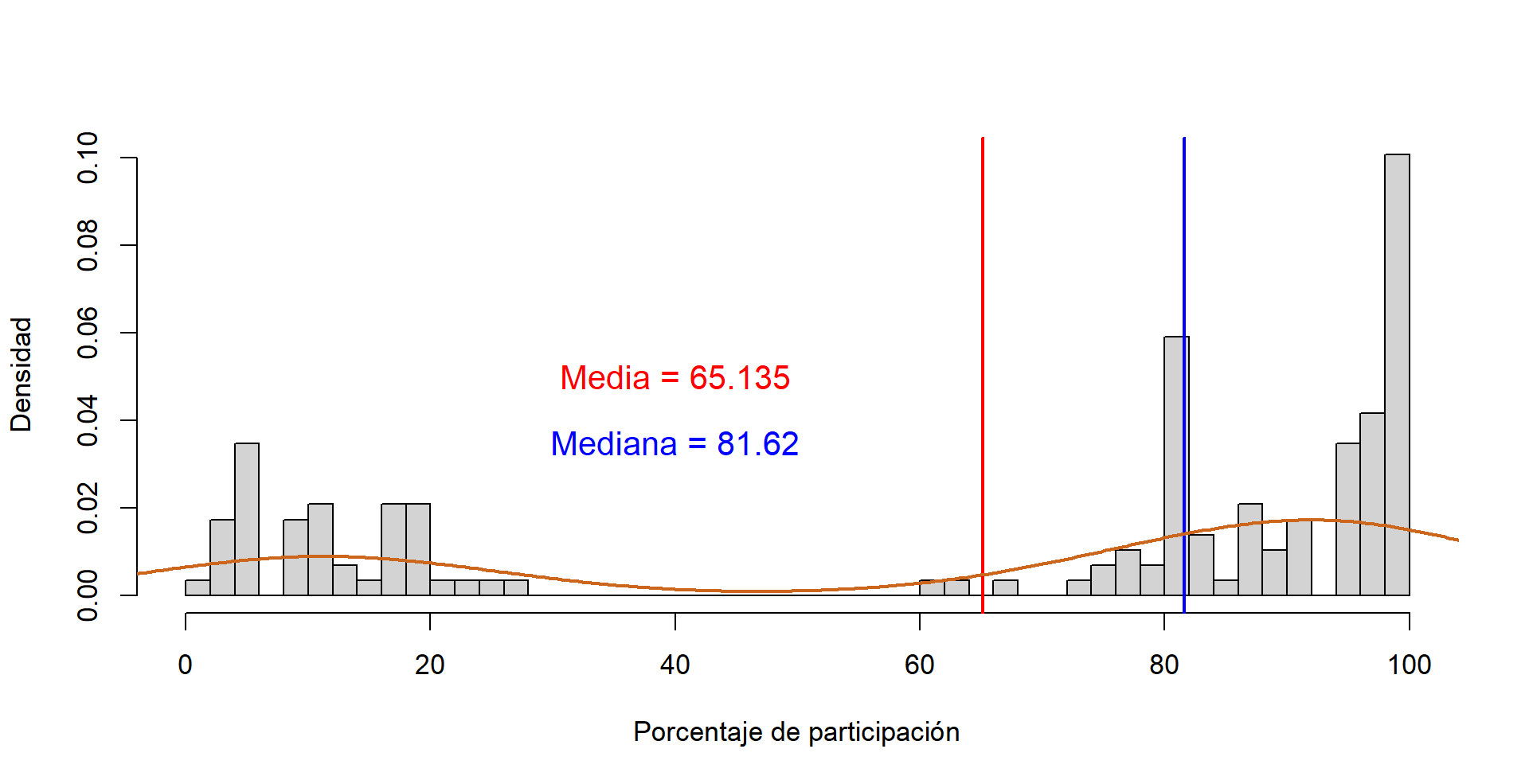}
    \caption{Porcentaje de participación de los senadores en las votaciones legislativas del Senado de la República 2006-2010.}
    \label{histpart}
\end{figure}

En la Figura \ref{histpart} se observa que la participación de los senadores fue desigual, dado que hubo unos senadores con tasas muy bajas de votación mientras que hubo otros con tasas muy altas. Las bajas tasas de participación son consecuencia de que varios senadores estuvieron por un periodo muy corto de tiempo en el congreso puesto que renunciaron, fueron sustituidos, reemplazos o fallecieron.

\begin{figure}[!h]
    \centering
    \includegraphics[scale=0.62]{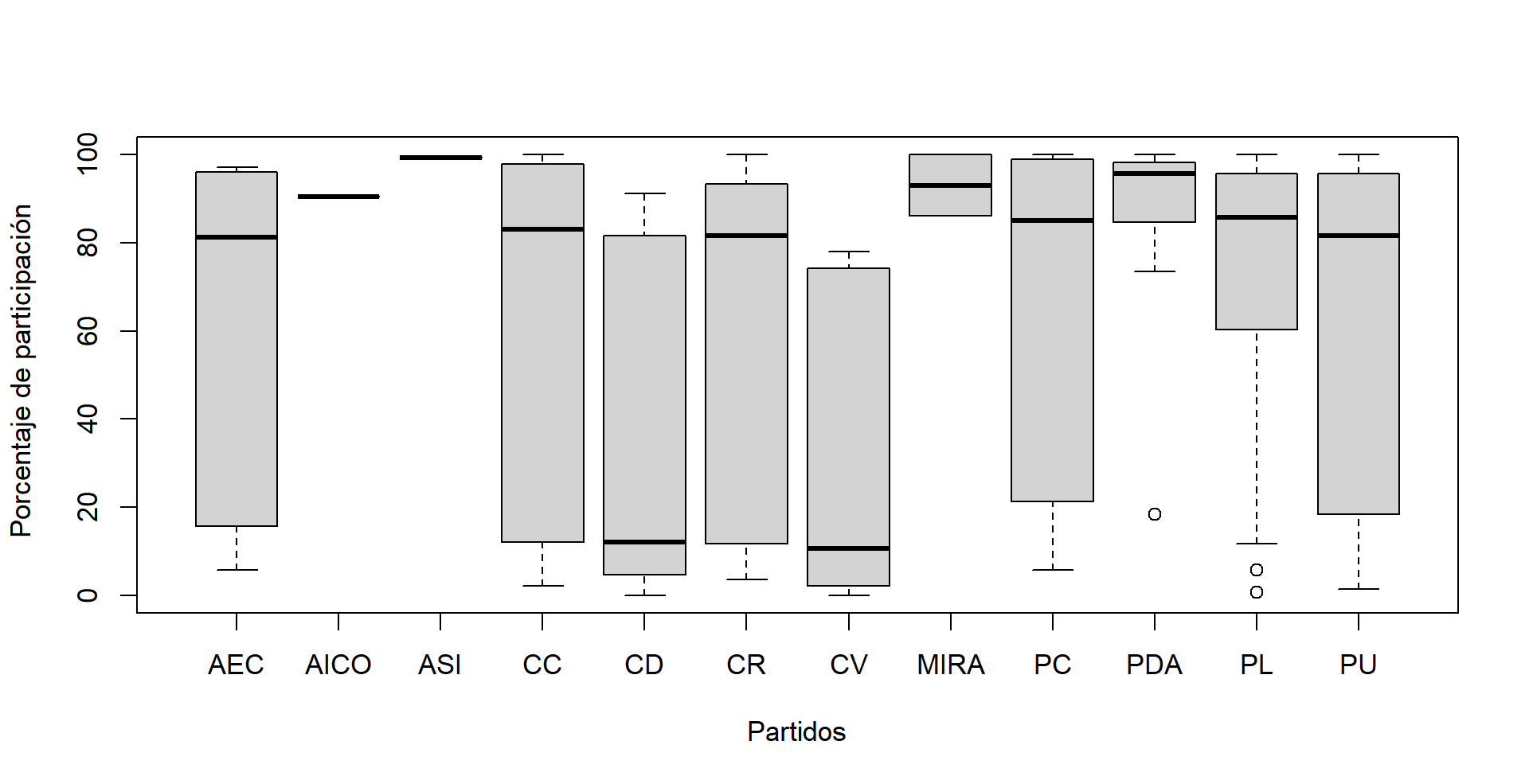}
    \caption{Porcentaje de participación por partido en las votaciones legislativas del Senado de la República 2006-2010.}
    \label{boxpart}
\end{figure}

En la Figura \ref{boxpart} se evidencia que la participación de los partidos de la coalición de Gobierno fue muy variable, lo cual se debe a que estos partidos fueron los más afectados por las investigaciones de la parapolítica. El PL también presenta una alta variabilidad, aunque menor a la de los partidos de coalición debido a que fueron pocos los senadores de esta colectividad que tuvieron lios con el escándalo de la parapolítica. De otra parte, la participación de los demás partidos fue más estable y mayoritariamente superior al $80\%$.

\begin{figure}[!t]
    \centering
    \includegraphics[scale=0.62]{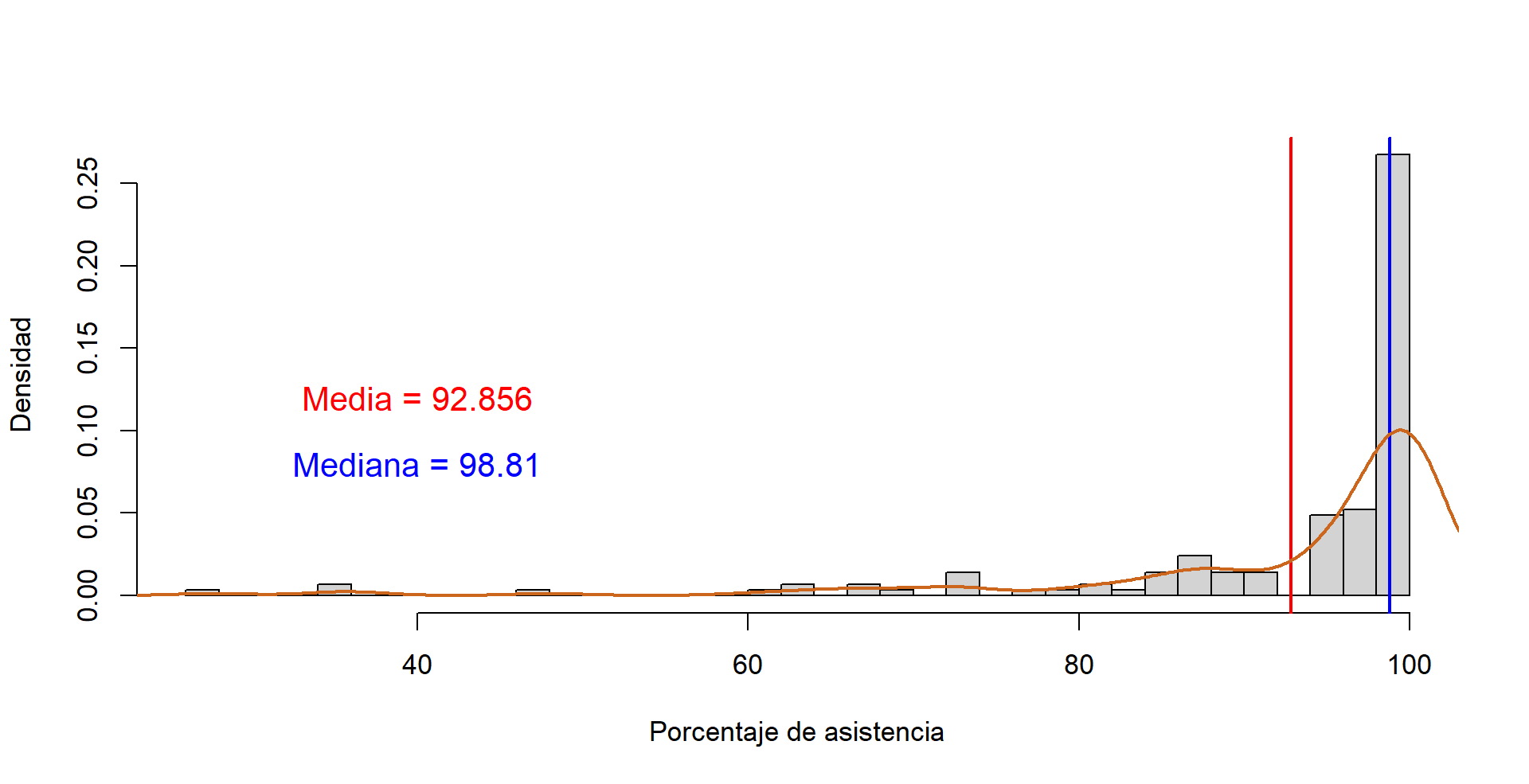}
    \caption{Porcentaje de asistencia de los senadores en las votaciones legislativas del Senado de la República 2006-2010.}
    \label{histasis}
\end{figure}

En la Figura \ref{histasis} se observa que la asistencia de los senadores a las votaciones legislativas en las cuales estaban activos fue en su mayoría alta, un $50\%$ de los senadores asistieron a más del $98\%$ de las votaciones en las cuales aparecían en lista. 

\begin{figure}[!b]
    \centering
    \includegraphics[scale=0.62]{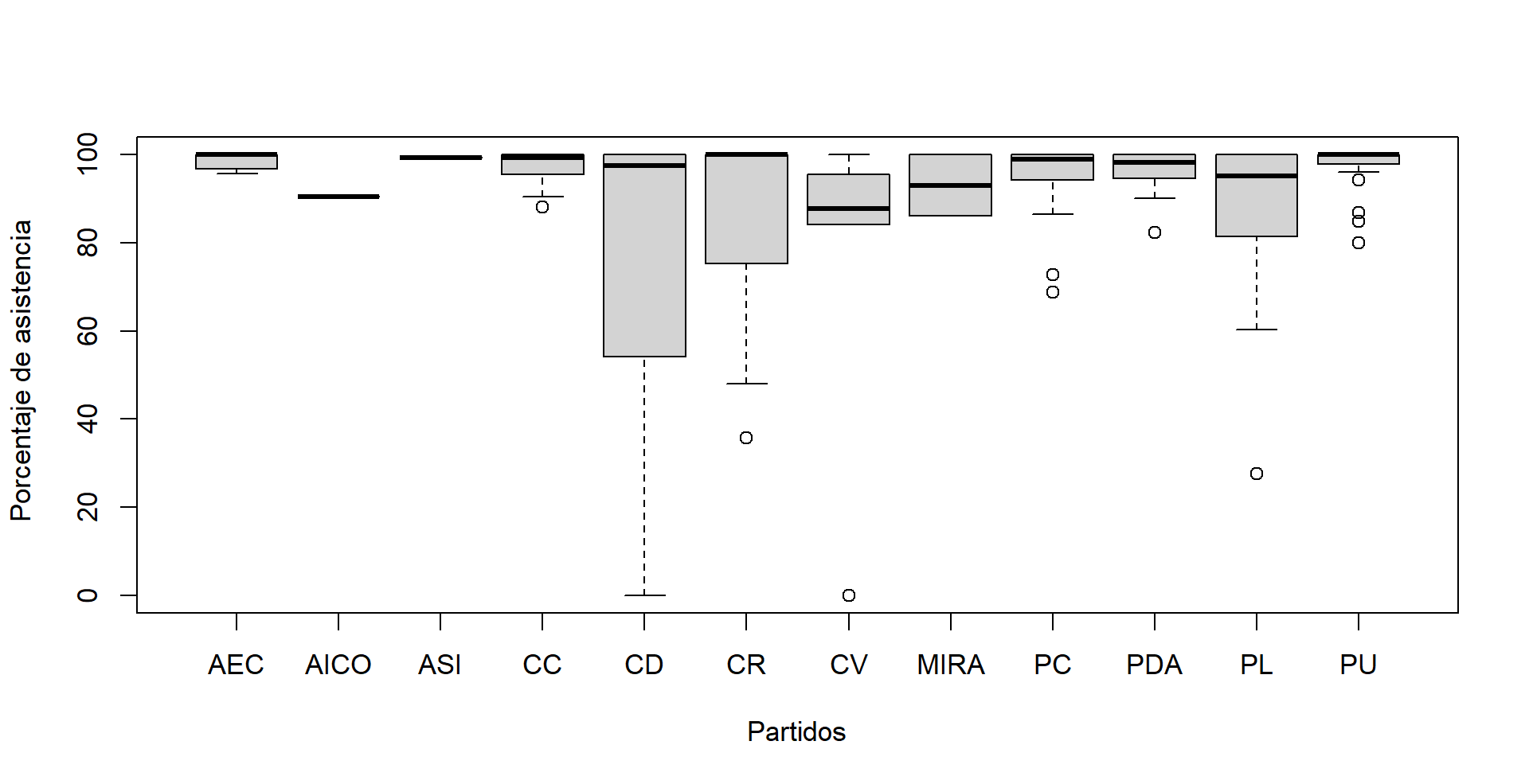}
    \caption{Porcentaje de asistencia por partido en las votaciones legislativas del Senado de la República 2006-2010.}
    \label{boxasist}
\end{figure}

Además, en la Figura \ref{boxasist} se aprecia que la asistencia por partidos fue más variable en los partidos CD y CR. En los demás partidos la asistencia fue mayoritariamente superior al $80\%$.

\begin{figure}[!t]
    \centering
    \includegraphics[scale=0.62]{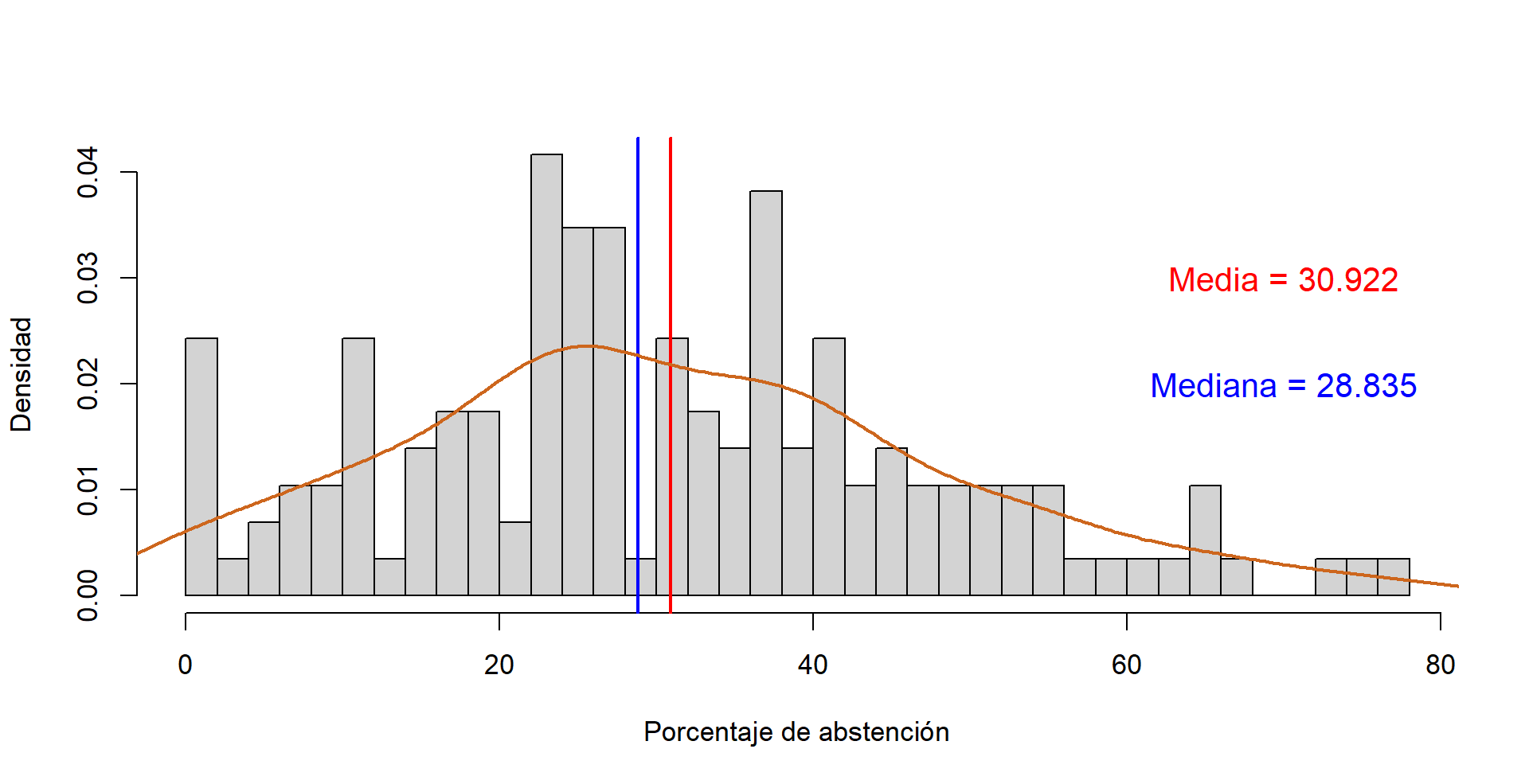}
    \caption{Porcentaje de abstención de los senadores en las votaciones legislativas del Senado de la República 2006-2010.}
    \label{histabst}
\end{figure}

En la Figura \ref{histabst} se contempla que la abstención de los senadores en las votaciones legislativas en las cuales participaron fue muy variable, aunque los valores de abstención más frecuentes son bajos-medios. El $72\%$ de los senadores tienen un porcentaje de abstención menor o igual al $40\%$. Solo el $7\%$ de los senadores presentan un porcentaje de abstención superior o igual al $60\%$.

\begin{figure}[!b]
    \centering
    \includegraphics[scale=0.62]{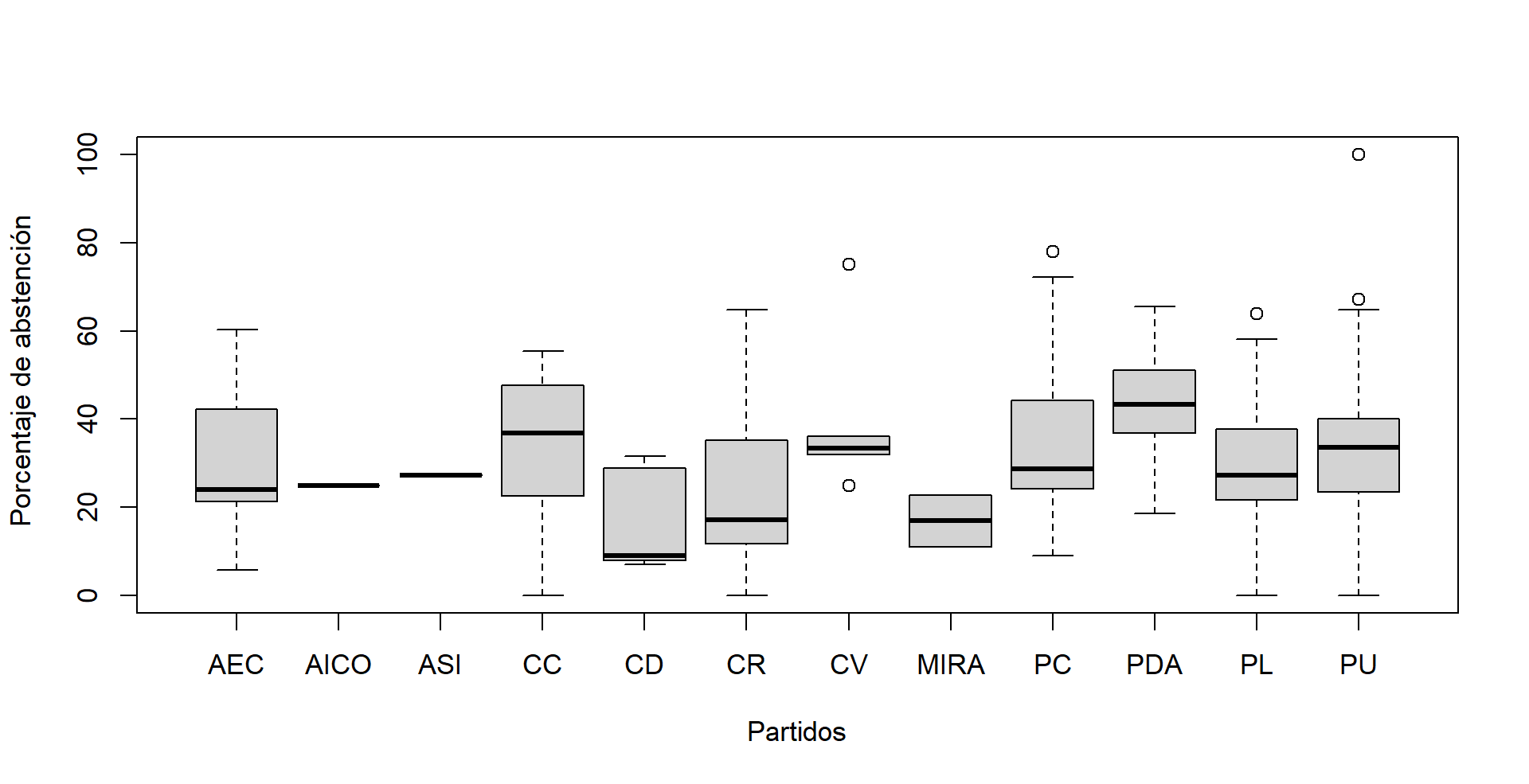}
    \caption{Porcentaje de abstención por partido en las votaciones legislativas del Senado de la República 2006-2010.}
    \label{boxabst}
\end{figure}

En la Figura \ref{boxabst} se aprecia que la mayoría de partidos tienen una abstención inferior al $60\%$. El partido con menor variabilidad en sus porcentajes de abstención es el MIRA, el cual a su vez es uno de los partidos con porcentajes de abstención más bajos. El partido de oposición PDA es el que presenta los mayores porcentajes de abstención.

\subsection{Parapolítica}

Parapolítica es el término con el que se identifica al escándalo político que se inició en el año 2006 que consistió en la alianza entre grupos paramilitares y algunos políticos con el fin de obtener cargos de elección popular tanto a nivel regional como nacional. Este escándalo se dio a conocer mediante las investigaciones realizadas por organismos judiciales del Estado, medios de comunicación y algunos miembros de partidos de oposición tales como el exsenador y actual Presidente de la República Gustavo Petro Urrego. El entonces representante a la cámara Gustavo Petro llevó a cabo varias denuncias de vínculos entre algunos congresistas, alcaldes y gobernadores con organizaciones paramilitares. Estas denuncias por parte de Gustavo Petro continuaron luego de que fuese elegido como senador para el periodo 2006-2010. Debido a esta labor Gustavo Petro ganó relevancia en el panorama político del país y se convirtió en una de las figuras más relevantes de la izquierda colombiana.

\begin{figure}[h]
    \centering
    \includegraphics[scale=0.57]{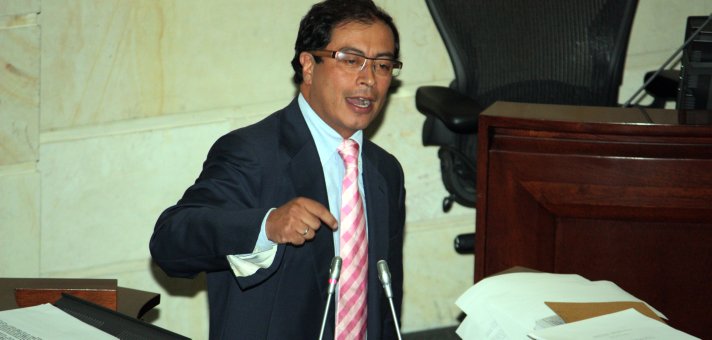}
    \caption{Gustavo Petro en un debate del Senado de la República 2006-2010\protect\footnotemark.}
    \label{petro}
\end{figure} 
\footnotetext{Foto: \url{https://www.lasillavacia.com/}.}

El escándalo de la parapolítica se dio a conocer luego del proceso de paz realizado por el expresidente Alvaro Uribe para lograr la desmovilización de las autodenominadas autodefensas unidas de Colombia (AUC). Los testimonios y las pruebas documentales entregadas por algunos de los exjefes paramilitares permitieron establecer las conexiones que habían entre políticos y paramilitares.

\newpage

De los partidos y movimientos políticos presentes en el congreso del 2006-2010 la mayoría tuvieron congresistas involucrados con la parapolítica. La mayoría de congresistas involucrados eran parte de partidos y movimientos de la coalición uribista. Entre estos partidos estaba el partido CD, fundado por el entonces senador Mario Uribe, primo del expresidente Alvaro Uribe, el cual tuvo la particularidad de que todos sus senadores electos fueron salpicados por la parapolítica. Uno de los casos más sonados fue el del exsenador Mario Uribe, esto debido a su parentesco con el entonces presidente Alvaro Uribe y al hecho de que era el jefe del partido CD. El exsenador Mario Uribe en el año 2011 fue condenado por parapolítica a 90 meses de prisión.

\begin{figure}[htb]
    \centering
    \includegraphics[scale=0.57]{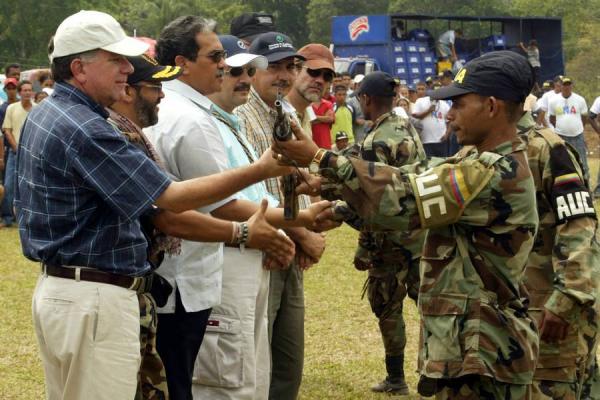}
    \caption{Proceso de Desmovilización de los paramilitares durante el Gobierno de Alvaro Uribe\protect\footnotemark.}
\end{figure} 
\footnotetext{Foto: \url{https://www.vanguardia.com/}.}

Al menos una tercera parte de los senadores del periodo 2006-2010 fueron salpicados por el escándalo de la parapolítica, muchos de ellos renunciaron al senado para no tener que ser investigados por la Corte Suprema de Justicia, otros continuaron siendo parte del Senado y hubo quienes fueron capturados en el transcurso de ese periodo. Algunos de los senadores de este periodo fueron vinculados a la parapolítica años después dado que las investigaciones por parapolítica siguen vigentes hoy en día y en los últimos años algunas de estas han pasado a la jurisdicción especial para la paz (JEP). Es importante resaltar que varias de las personas que estuvieron involucrados con la parapolítica fueron absueltos debido a que no se encontraron pruebas en su contra o a que se comprobó que no tuvieron vínculos con paramilitares.

Un conjunto de 12 movimientos políticos tuvieron representación en el Senado de la República 2006-2010, 8 de los cuales tuvieron senadores involucrados con la parapolítica. En la Tabla \ref{tab:table1} se observa el porcentaje de senadores por partido involucrados en el escándalo de la parapolítica, donde solo están los partidos que tuvieron senadores vinculados con la parapolítica. Los movimientos políticos con menor porcentaje de senadores involucrados son el PC, el PL y el PU con 25$\%$, 27.3$\%$ y 37.9$\%$, respectivamente. Los movimientos políticos con un mayor porcentaje de senadores involucrados son CV, AEC y CD con 66.7$\%$, 50$\%$ y 50$\%$, respectivamente.

El PU en ese entonces era el partido de Gobierno. Este partido fue fundado por Juan Manuel Santos con el objetivo de agrupar a muchos de los políticos simpatizantes del expresidente Uribe y de esta forma poder acompañar al Gobierno en sus iniciativas legislativas. El PC y CR hacían parte de la coalición de Gobierno y junto con el PU eran los que mayor visibilidad y tamaño tenían entre los movimientos políticos afines al expresidente Uribe. Debido al escándalo de la parapolítica varios candidatos al Senado por estas tres colectividades fueron expulsados de sus respectivos partidos, esto produjo que muchos de estos políticos buscaran aval en otras colectividades tales como AEC, CV, CD, y CC.

\begin{table}[h]
	\centering
\begin{tabular}{|c|c|c|c|c|c|c|c|}
\hline
AEC & CC & CD & CR & CV & PC & PU & PL \\ \hline
50$\%$ & 45.5$\%$ & 50$\%$ & 43.5$\%$ & 66.7$\%$ & 25$\%$ & 37.9$\%$ & 27.3$\%$  \\ \hline
\end{tabular}
\caption{Porcentaje de senadores por movimiento político involucrados en el escándalo de la parapolítica.}
\label{tab:table1}
\end{table}

El Movimiento AEC surge a finales del año 2005 como unión del partido ALAS y del Movimiento Equipo Colombia, esto con el fin de sobrevivir al umbral\footnote{La reforma política de 2003 impuso un umbral del $2\%$ de los votos válidos en las elecciones legislativas para que los partidos puedan mantener su personeria jurídica.} de las elecciones legislativas. El Partido Alas fue fundado por Álvaro Araujo Castro, quien fue elegido como senador en el año 2006 y tiempo después fue condenado por parapolítica. El Movimiento Equipo Colombia fue fundado por Luis Alfredo Ramos en los años ochenta. Luis Alfredo Ramos ocupó varios cargos de elección popular entre los cuales resalta el de Gobernador de Antioquía, fue condenado por parapolítica en el año 2021.

El Movimiento CV fue fundado en el año 2003 como consecuencia de la reforma política de ese mismo año que fomentaba la unión de los partidos más pequeños para sobrevivir al umbral impuesto para las elecciones legislativas. Entre sus candidatos al senado para el periodo 2006-2010 figuraban Habib Merheg, Vicente Blel y Dieb Maloof. Los dos primeros habían sido rechazados por el PU debido a presuntos vínculos con grupos paramilitares, mientras que el tercero había sido expulsado por el PL debido a presuntos vínculos con el narcotráfico. Los tres fueron parte del Senado del periodo 2006-2010 y a su vez terminaron siendo investigados por parapolítica.

El Partido CC surgió a finales del siglo 20 y para las elecciones legislativas del 2006 recibió a varios políticos que habían sido expulsados de otros partidos por presunto vínculos con paramilitares. Durante el periodo 2006-2010 fue parte de la coalición de Gobierno y a su vez uno de los partidos más golpeados por el escándalo de la parapolítica.

\begin{figure}[!h]
    \centering
    \includegraphics[scale=0.6]{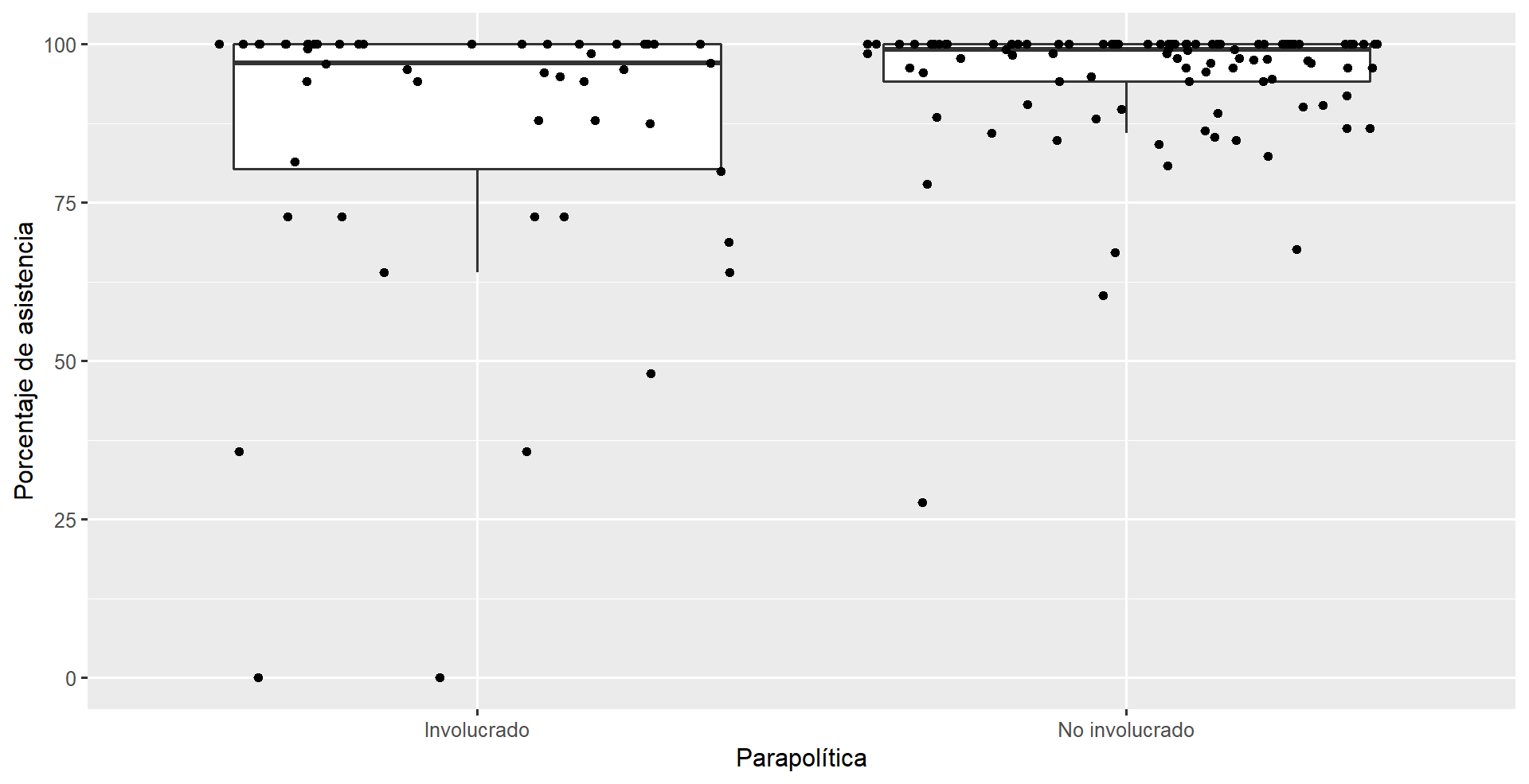}
    \caption{Porcentaje de asistencia para involucrados y no involucrados en parapolítica del Senado de la República 2006-2010.}
    \label{boxasispar}
\end{figure}

En la Figura \ref{boxasispar} se aprecia que estar involucrado en el escándalo de la parapolítica tiene un efecto considerable en el porcentaje de asistencia. La distribución del porcentaje de asistencia de los involucrados en parapolítica es menos concentrada que la de los no involucrados, y a su vez los senadores involucrados tienden a tener porcentajes más bajos de asistencia. Ambas distribuciones tienen asimetría negativa, igual que la distribución global del porcentaje de asistencia (ver Figura \ref{histasis}).

\begin{figure}[!t]
    \centering
    \includegraphics[scale=0.6]{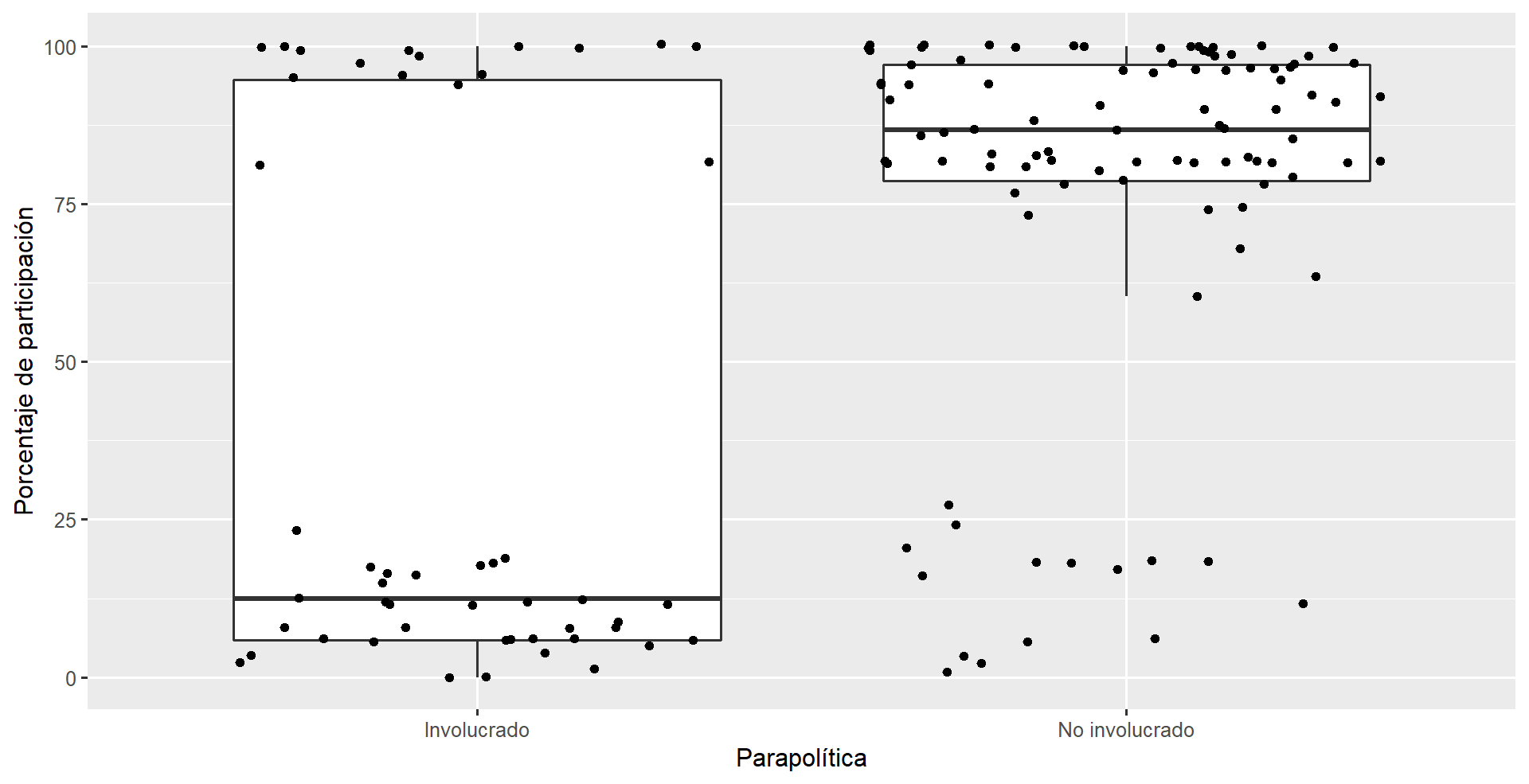}
    \caption{Porcentaje de participación para involucrados y no involucrados en parapolítica del Senado de la República 2006-2010.}
    \label{boxpartpar}
\end{figure}

En cuanto al porcentaje de participación de los involucrados en parapolítica, como era de esperarse, es muy inferior al porcentaje de participación de los no involucrados, tal como se evidencia en la Figura \ref{boxpartpar}. Esto está en concordancia con el hecho de que la principal y mayoritaria causa de renuncia al Senado durante el periodo 2006-2010 fue el escándalo de la parapolítica. La distribución del porcentaje de participación de los no involucrados luce simétrica, con una mediana superior al 80$\%$. En cambio, la distribución del porcentaje de participación de los involucrados tiene asimetría negativa con una mediana inferior al $20\%$. Esto último evidencia que una buena parte (más del $50\%$) de los involucrados en el escándalo de la parapolítica tuvieron una estancia muy corta en el Senado de ese periodo.

\begin{figure}[!b]
    \centering
    \includegraphics[scale=0.6]{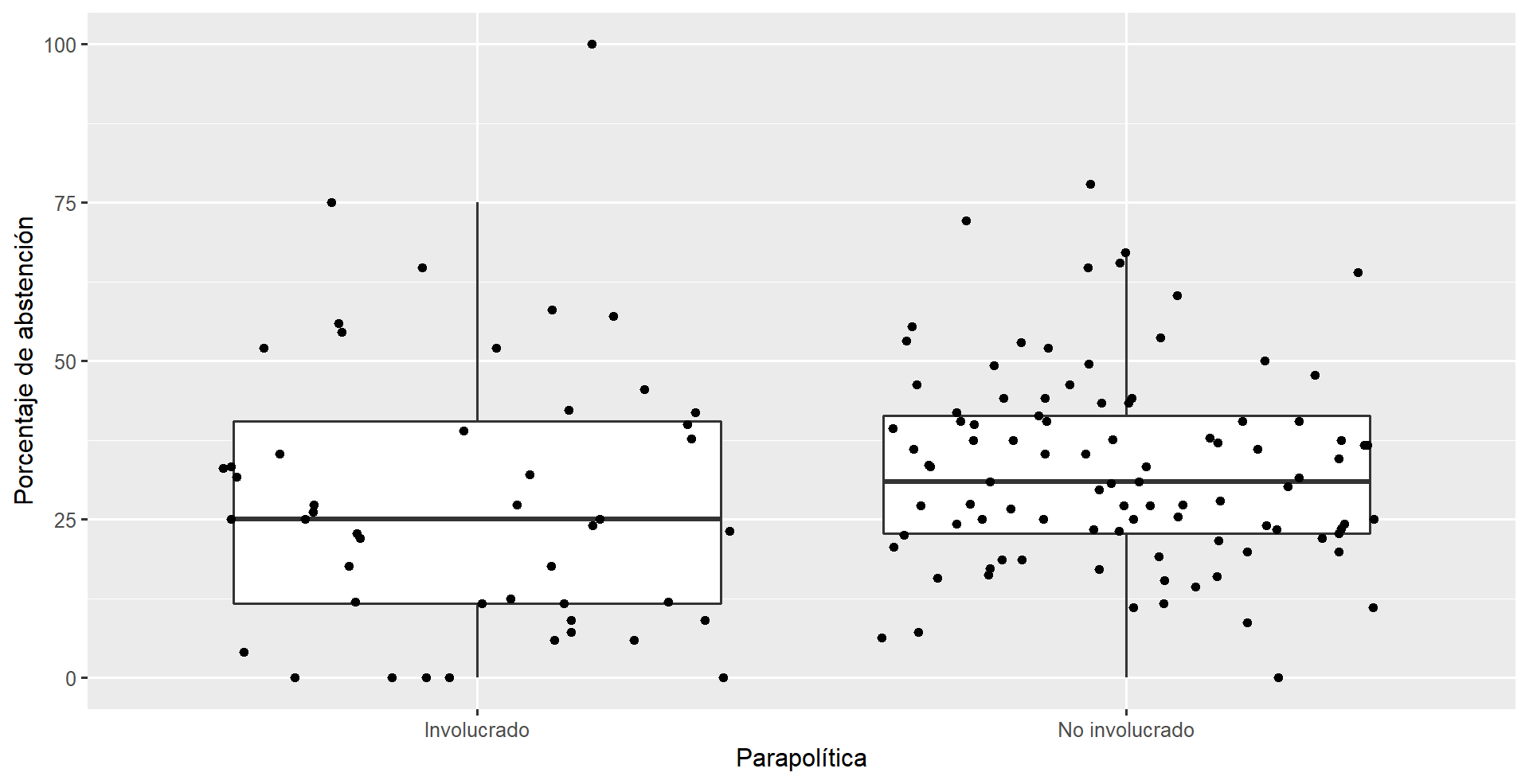}
    \caption{Porcentaje de abstención para involucrados y no involucrados en parapolítica del Senado de la República 2006-2010.}
    \label{boxabstpar}
\end{figure}

En la Figura \ref{boxabstpar} se contempla que el porcentaje de abstención para los involucrados y los no involucrados no tiene diferencias sustanciales. En ambos casos la abstención se ubica mayoritariamente en un rango entre 0$\%$ y $40\%$ con una mediana ligeramente superior para el grupo de los no involucrados. Aunque se podría pensar que el estar bajo investigación por el escándalo de la parapolítica habría hecho que el porcentaje de abstención fuera superior, dado que posiblemente los senadores involucrados no quisieran llamar la atención, esto no se refleja con los datos observados. La realidad es que muchos de los senadores involucrados renunciaban tan pronto se conocía que la Corte Suprema de Justicia los investigaría. Los senadores involucrados renunciaban al Senado con el propósito de que la investigación pasara a la Fiscalía General de la Nación, ya que la Corte Suprema de Justicia para esa época produjo varias capturas y condenas a corto plazo. Eso posiblemente explica que no se evidencien diferencias considerables entre los porcentajes de abstención de los involucrados y los no involucrados.

\section{Computación}

La dimensión del espacio político es 1. En este caso, los parámetros $\boldsymbol{\alpha}_j$, $j=1,\ldots,m$, y $\boldsymbol{\beta}_i$, $i=1,\ldots,n$ son escalares, donde el número de listas de votación plenaria es $m$ y el número de senadores es $n$. Debido a que cualquier traslación o rotación de los puntos ideales y propuestas mantiene constantes las distancias entre los puntos ideales y las alternativas de votación, es necesario fijar la posición de 2 senadores, los cuales se les conoce como senadores de anclaje \cite{clinton2004statistical}. En este caso solo se necesitan dos senadores de anclaje porque la dimensión es 1. Lo usual es tomar dos senadores de diferentes orillas políticas como senadores de anclaje, uno de ellos con posición $1$ y el otro $-1$.

Para el Senado 2006-2010 se establecen dos senadores de anclaje de dos partidos ideológicamente opuestos. Se fijan las posiciones del senador Guillermo Alfonso Jaramillo Martínez del PDA y del senador Carlos Cárdenas Ortiz del PU en -1 y 1, respectivamente. Los senadores Jorge de Jesús Castro Pacheco de CV, Alvaro Alfonso García Romero de CD y Jairo Enrique Merlano Fernandez del PU no hacen parte del modelo debido a que no hay registros de votación de ellos a favor o en contra en ninguna lista de plenaria. Por lo tanto, el número de senadores que se usan para el modelo es $n=144$.

Para los hiperparámetros se usan los mismos valores que en \cite{luque2022bayesian} y \cite{luque2021metodos}. Así $\alpha_0=0$ y $A_0=25$, $b_i=0$ y $B_i=1$. Esta elección de los hiperparámetros es muy conveniente dado que esta elección implica distribuciones previas no informativas para los parámetros $\mu_j$ y $\boldsymbol{\alpha}_j$, y su vez garantiza que a priori más del $99\%$ de los puntos ideales se encuentran entre $-3$ y $3$ \citep{luque2021metodos}.

Se usa como función de enlace la función logit. Otra opción para la función de enlace es la función probit, la cual produce resultados similares \cite{luque2022bayesian}. Se implementa un algoritmo de Cadenas de Markov de Monte Carlo (MCMC, por sus siglas en inglés; ver por ejemplo, \citealt{gamerman2006markov} y \citealt{gelman2013bayesian}) para obtener las muestras de la distribución posterior $\textsf{p}(\mu,\boldsymbol{A},\boldsymbol{B}\mid \boldsymbol{Y})$ la cual se encuentra inmersa en un espacio de $2m+n-2=414$ dimensiones.

\begin{figure}[!t]
    \centering
    \includegraphics[width=12.5 cm,height=18 cm]{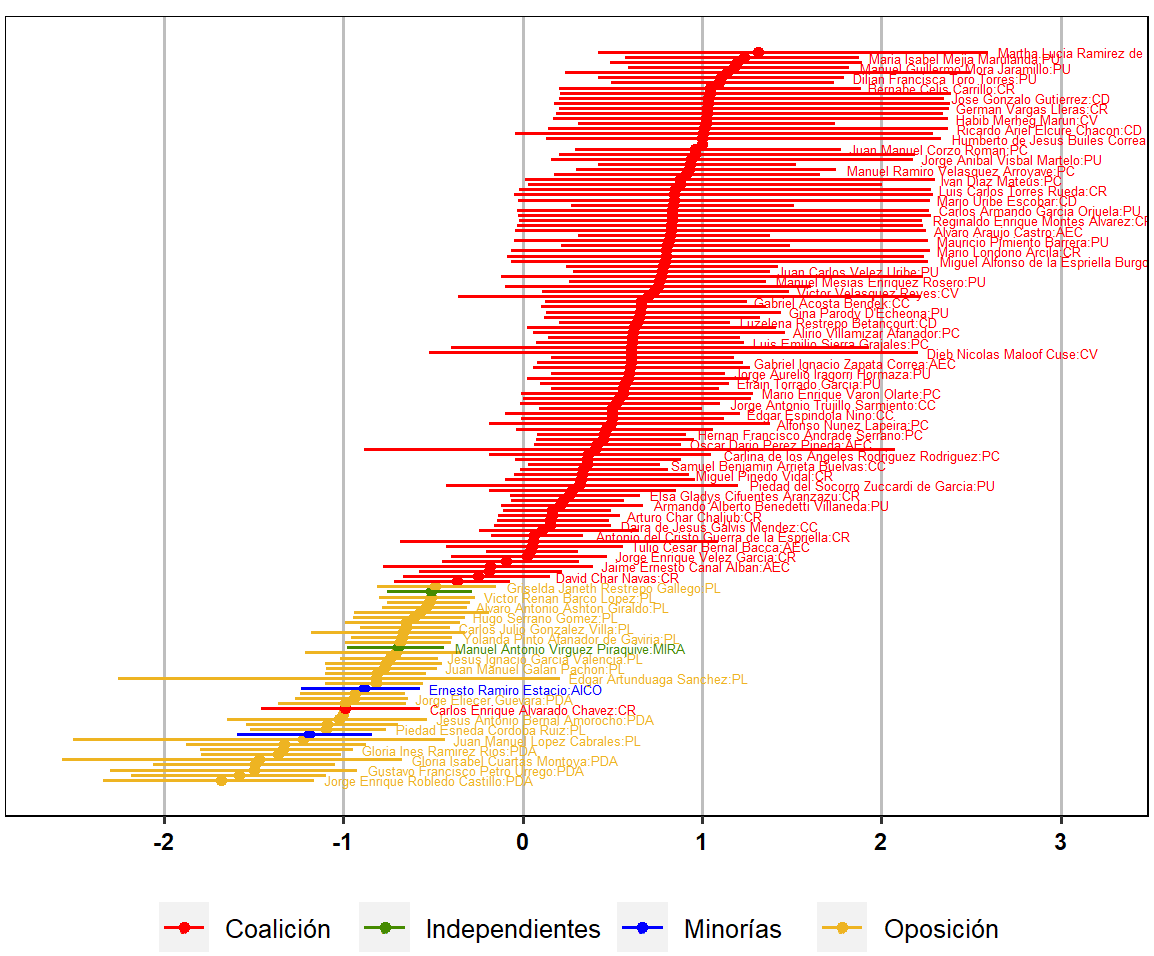}
    \caption{Puntos ideales estimados bajo la elección de legisladores de anclaje opuestos y a diferentes distancias del centro. Los puntos representan la estimación puntual y las líneas horizontales representan el intervalo de credibilidad simétrico basado en percentiles al $95\%$.}
    \label{credibilidad}
\end{figure}

El modelo se implementa en R en su versión version 4.2.1 (\url{https://www.R-project.org/}) mediante el uso de STAN en su versión 2.26.1 (\url{https://mc-stan.org}). Se usa una adaptación del código\footnote{En el siguiente link se puede acceder al conjunto de datos y al código del modelo: \url{https://drive.google.com/drive/folders/18b5M-SOc22kcXhF1DY4123rXFMsHWVRV?usp=sharing}.} desarrollado en \cite{luque2022bayesian} y \cite{luque2021metodos} (disponible en \url{https://github.com/Luque-ZabalaC}). Se realizan 80000 iteraciones, donde las primeras 16000 corresponden al periodo de calentamiento. Los valores iniciales de los puntos ideales se toman como $1$ si el senador es parte de la coalición y $-1$ en caso contrario. Con el fin de disminuir la influencia de los valores iniciales de los puntos ideales y la autocorrelación entre las muestras se reduce el tamaño de la cadena usando un muestreo sistemático, en el cual se toma 1 muestra cada 5 iteraciones. Por lo tanto, las inferencias posteriores se llevan a cabo usando 12800 (64000/5) iteraciones.
Para la convergencia del modelo ver anexo.

\section{Puntos ideales}

A continuación se muestran los resultados producto de la implementación del estimador de punto ideal Bayesiano estándar para el Senado de la República de Colombia Periodo 2006-2010. El análisis que se presenta está encaminado a distinguir rasgos latentes, patrones partidistas y de coalición que se encuentran en las decisiones legislativas de los senadores. Además, se analizan los puntos ideales de los senadores teniendo en cuenta si han estado o no involucrados en el escándalo de la parapolítica.

En la Figura \ref{credibilidad} se visualiza los puntos ideales junto con sus respectivo intervalos de credibilidad de los 144 senadores que hacen parte del análisis. Los únicos puntos que no tienen intervalo de crebilidad son los de los senadores de anclaje debido que sus posiciones están fijadas de antemano, luego no se tienen que estimar y por lo tanto no tienen incertidumbre asociada. Se observan patrones en el comportamiento de estos puntos según el grupo político al cual pertenecen los senadores durante el periodo de interés. Los senadores de oposición se ubican al lado izquierdo del espacio político en contraposición a los senadores que hacen parte de la coalición de gobierno\footnote{La noción de izquierda y de derecha es artificial, la cual es resultado de la elección de las posiciones de los senadores de anclaje. es suficiente notar que una reflexión del espacio cambia las posiciones pero deja invariante las contraposiciones encontradas.}. Todos los senadores de los grupos de oposición, independientes y minorías se ubican a la izquierda del 0. La mayoría ($94.3\%$) de los senadores de la coalición se ubican a la derecha de 0.

La media posterior de los puntos ideales de los senadores se ubica en el intervalo $(-1.68,1.31)$. En la Tabla \ref{tab:table2} se encuentra el mínimo, el máximo y el coeficiente de variación a posteriori de los puntos ideales por grupo político. A partir de estas medidas se evidencia que los independientes y las minorías son los grupos más estables en términos de sus preferencias, a diferencia de los de la coalición y los de la oposición, que corresponden a los grupos más heterogéneos. Lo anterior es contrario a lo que sucede en el Senado 2010-2014, en donde los grupos más estables son la oposición y la coalición \citep{luque2021metodos}. Al comparar los cuatro grupos se identifica que el grupo de la coalición de gobierno es el más disperso respecto a los puntos ideales de sus miembros.

\begin{table}[H]
	\centering
\begin{tabular}{|c|c|c|c|c|}
\hline
 & Coalición & Independientes & Minorías & Oposición \\ \hline
Mínimo & -0.99 & -0.70 & -1.19 & -1.68 \\ \hline
Máximo & 1.31 & -0.51 & -0.89 & -0.49 \\ \hline
CV($\%$) & 63.95 & 21.71 & 20.80 & 37.07   \\ \hline
\end{tabular}
\caption{Medidas de resumen de los puntos ideales estimados por grupo político. Para cada grupo se muestra el mínimo, el máximo y el coeficiente de variación (CV).}
\label{tab:table2}
\end{table}

En la Tabla \ref{tab:table3} se encuentra el mínimo, el máximo y el coeficiente de variación a posteriori de los puntos ideales por partido político. Con base en estas mediciones se evidencia que el MIRA (el partido independiente), y el PDA y el PL (los partidos de oposición) son los partidos con menor variabilidad. Entre los partidos de la coalición hay tres con una variabilidad mucho más alta que el resto, a saber, CR, AEC y CC. El Partido CR fue fundado en 1997 por exmiembros del PL. Debido a esto, algunos de sus senadores habían sido parte del PL, el cual para el periodo en cuestión figuraba como partido de oposición. La presencia de senadores en CR con pasado en el PL, que para ese cuatrenio figuraba como partido de oposición, podría explicar la heterogeneidad sustancial entre sus senadores. El Partido AEC fue fundado por Alvaro Araujo Castro y Luis Alfredo Ramos, dos personajes con orígenes políticos distintos. Alvaro Araujo Castro lideraba al Movimiento Alas, que era un grupo al interior del PL, hasta que decidieron separarse del liberalismo para apoyar al expresidente Uribe. Por otro lado, Luis Alfredo Ramos lideraba al Movimiento Equipo Colombia, que era un grupo al interior del PC, hasta que en 2004 decidieron separarse de ese partido \citep{castro2011transformaciones}. La variabilidad entre los senadores de AEC podría ser causada por la fusión de dos movimientos que eran parte de partidos tradicionalmente opuestos. La variabilidad de los miembros de CC podría deberse a los cambios de partido que se dieron luego de la reforma política de 2009. Algunos de los senadores de CC hicieron cambio de partido, como por ejemplo el senador Carlos Emiro Barriga Peñaranda, quién se pasó al PC, o la senadora Daira de Jesús Galvis Méndez, quién se hizo miembro de CR.

\begin{table}[H]
	\centering
	\scalebox{0.8}{
\begin{tabular}{|c|c|c|c|c|c|c|c|c|c|c|c|c|}
\hline
& PDA & MIRA & AEC & CC & CD & CR & CV & PC & PU & PL & AICO & ASI \\ \hline
Mínimo & -1.68 & -0.70 & -0.18 & -0.37 & 0.46 & -0.99 & 0.41 & 0.34 & 0.06 & -1.23 & -0.89 & -1.19 \\ \hline
Máximo & -0.94 & -0.51 & 1.14 & 0.94 & 1.04 & 1.05 & 1.19 & 1.09 & 1.31 & -0.49 & -0.89 & -1.19 \\ \hline
CV ($\%$) & 18.75 & 21.71 & 99.89 & 82.42 & 30.59 & 145.86 & 40.12 & 29.89 & 39.11 & 25.75 & NA & NA \\ \hline
\end{tabular}}
\caption{Medidas de resumen de los puntos ideales estimados por partido político. Para cada partido se muestra el mínimo, el máximo y el coeficiente de variación (CV)\protect\footnotemark.}
\label{tab:table3}
\end{table}
\footnotetext{Los partidos AICO y ASI no tienen CV debido a que solo tuvieron un senador durante ese cuatrenio.}

Los intervalos de credibilidad más amplios (ver Figura \ref{credibilidad}) corresponden a los senadores que tuvieron una participación muy baja (por debajo del 10$\%$). Esto sucede con los senadores Vicente Blel de CV, Dieb Maloof de CV, Olano Portela del PU, Edgar Artunduaga del PL y Luis Guillermo Vélez, quiénes tuvieron porcentajes de participación de $2.21\%$, $5.15\%$, $3.68\%$, $0.74\%$ y $5.88\%$, respectivamente. Entre más amplio sea el intervalo de credibilidad de un punto ideal, mayor incertidumbre se tiene acerca de su estimación. Es natural que los intervalos más amplios estén asociados a los puntos ideales de los senadores con bajos niveles de participación, dado que entre menor sea la participación, mayor es la incertidumbre acerca del punto ideal correspondiente. En la Figura \ref{credibilidad} se observa que el ancho de los intervalos de credibilidad crece cuando los puntos ideales se alejan del centro. Además se observa que los intervalos de credibilidad suelen ser más amplios para los senadores miembros de la coalición, lo cual se debe a que la coalición agrupa a la mayoría de los senadores ($72\%$ de los senadores pertenecen a la coalición). Este comportamiento también se observa en el Senado 2010-2014 \citep{luque2022bayesian}.

\begin{figure}[!t]
 \centering
  \subfigure[Distribución marginal posterior de $\beta_2$]{\includegraphics[width=0.48\textwidth]{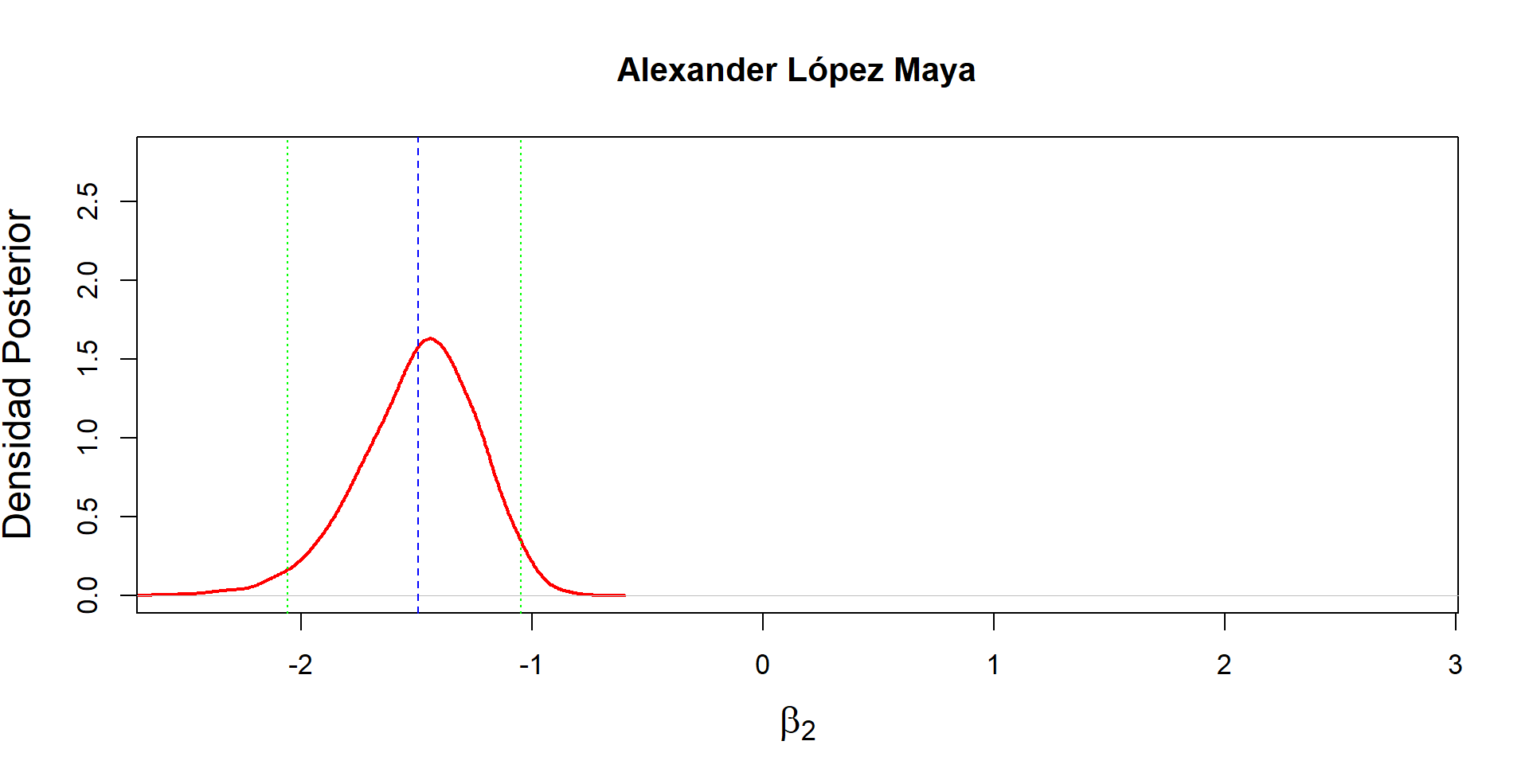}} \label{11}
  \subfigure[Distribución marginal posterior de $\beta_{22}$]{\includegraphics[width=0.48\textwidth]{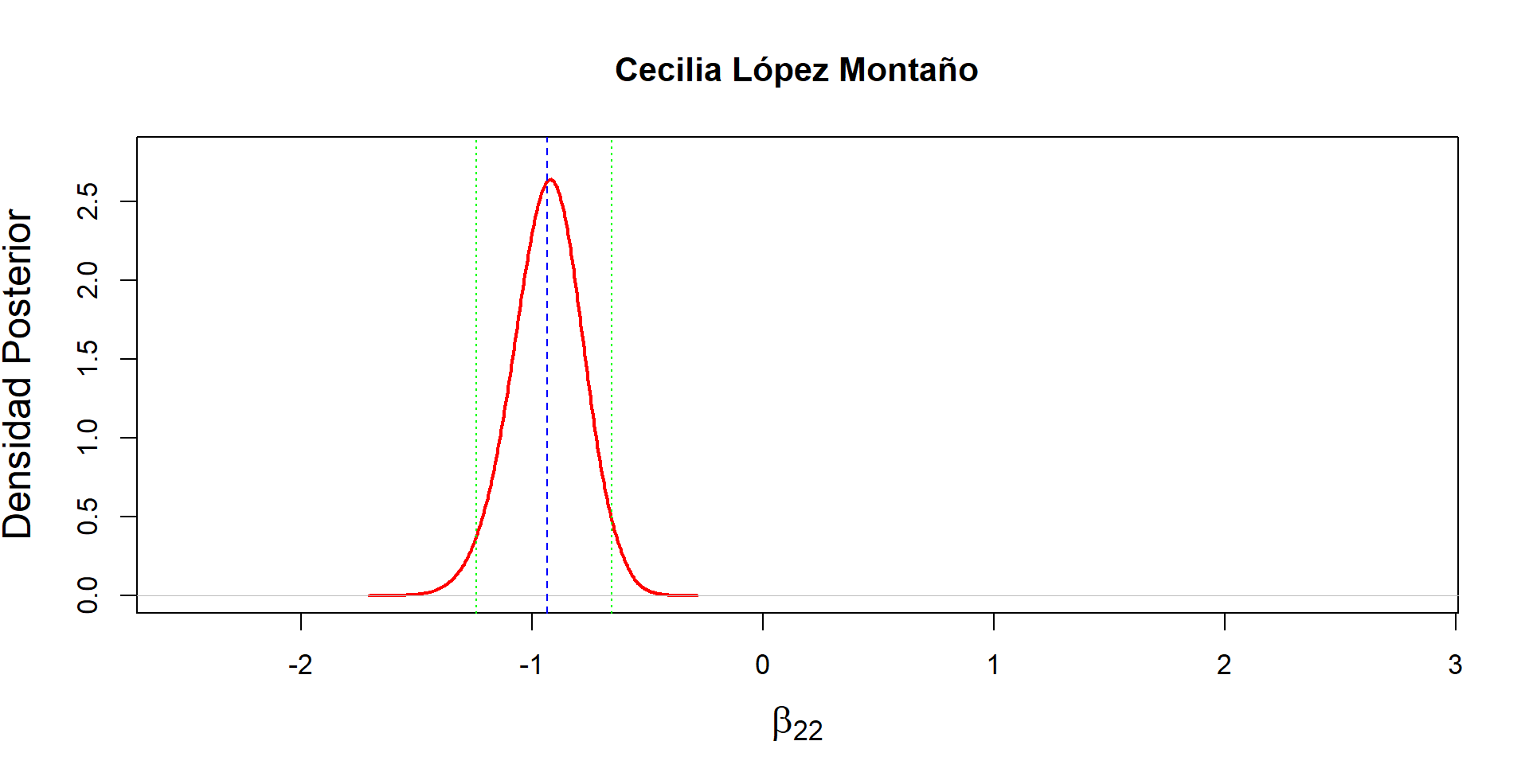}} \label{12} \\
  \subfigure[Distribución marginal posterior de $\beta_{110}$]{\includegraphics[width=0.48\textwidth]{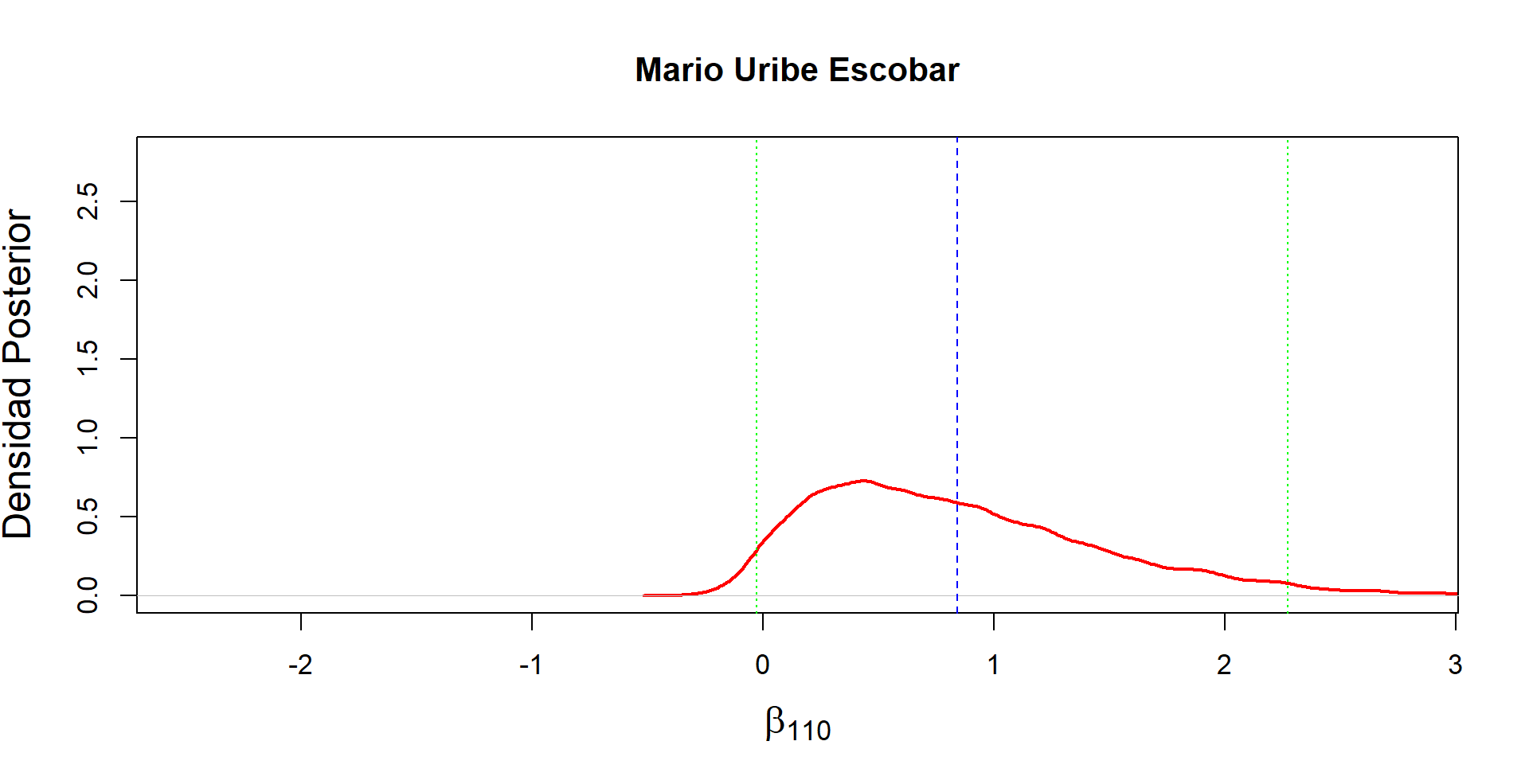}} \label{21}
  \subfigure[Distribución marginal posterior de $\beta_{142}$]{\includegraphics[width=0.48\textwidth]{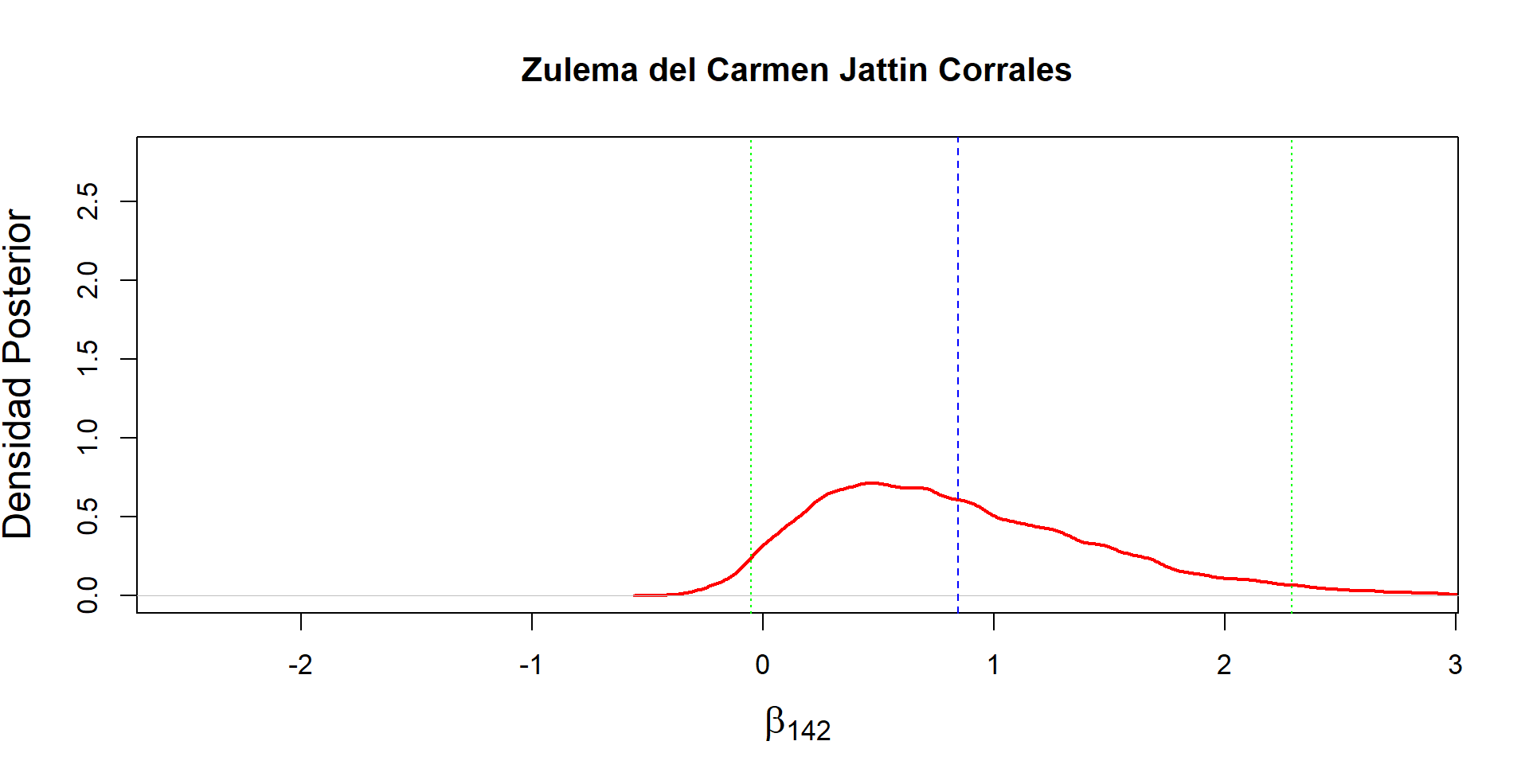}} \label{22}
 \caption{Distribución marginal posterior de los puntos ideales de los senadores Alexander López Maya del PDA ($\beta_1$), Cecilia López Montaño del PL ($\beta_22$), Mario Uribe Escobar de CD ($\beta_{110}$) y Zulema del Carmen Jattin Corrales del PU ($\beta_{142}$). La línea azul corresponde a la media posterior y las líneas verdes corresponden al intervalo de credibilidad al 95$\%$.} 
 \label{densidades}              
\end{figure}

De los 144 puntos ideales estimados hay 94 que son significativamente distintos de cero ($65.3\%$). Además, 49 de los 50 puntos ideales que no son distinguibles de cero ($98\%$), son de senadores de la coalición. El otro punto ideal no distinguible de cero es el del senador Edgar Artunduaga del PL, quién solo participó en una lista de votación de plenaria del periodo en cuestión. De estos 50 puntos ideales 16 corresponden a senadores de CR ($32\%$), partido que durante el periodo de interés tuvo 23 senadores. A su vez, los partidos AEC, CC y CD también presentan la característica de que al menos la mitad de sus senadores tienen puntos ideales no distinguibles del cero. Los senadores con puntos ideales distinguibles del cero tienen probabilidad más alta de votar a favor o en contra (es decir, probabilidad más alta de no abstenerse).

En la Figura \ref{densidades} se visualiza la distribución marginal posterior de los puntos ideales de los senadores Alexander López Maya del PDA ($\beta_{2}$), Cecilia López Montaño del PL ($\beta_{22}$), Mario Uribe Escobar de CD ($\beta_{110}$) y Zulema del Carmen Jattin Corrales del PU ($\beta_{142}$). A diferencia de los senadores Alexander López y Cecilia López, el senador Mario Uribe y la senadora Zulema Jattin estuvieron involucrados en el escándalo de la parapolítica, razón por la cual no estuvieron presentes en el Senado durante todo ese periodo. La distribución marginal que luce simétrica respecto a la media es la del punto ideal de la senadora Cecilia López Montaño del PL, las demás distribuciones presentan sesgo a izquierda o derecha. La distibución marginal del punto ideal del senador ALexander López presenta un sesgo mucho menor que la de los puntos ideales de los senadores Mario Uribe y Zulema Jattin. Las distribuciones no simétricas respecto a la media están asociadas a los senadores que tienen mayor incertidumbre en la estimación de su punto ideal. Para las distribuciones que se muestran en la Figura \ref{densidades} se observa que los puntos ideales de los senadores Mario Uribe Escobar y Zulema del Carmen Jattin Corrales son de los que menos certeza se tiene. Este patrón tambíen se observa en \citep{luque2022bayesian}.

\subsection{Probabilidades de estar en los extremos o en el centro}

Los senadores con más alta probabilidad de estar en los extremos del espacio político son los que hacen parte de la oposición o de la coalición (en \cite{luque2022bayesian} tambíen se observa el mismo comportamiento para el Senado 2010-2014). En particular, los 7 senadores que tienen más probabilidad de estar al extremo izquierdo son del PDA y 4 de los 5 que tienen más probabilidad de estar al extremo derecho son del PU. Los senadores con mayor probabilidad de tener un punto ideal menor que $-1$ son: Jorge Enrique Robledo Castillo del PDA ($99.6\%$), Jaime Dussan Calderón del PDA ($99.3\%$), Alexander López Maya del PDA ($98.8\%$), Luis Carlos Avellaneda Tarazona del PDA ($97.8\%$), Gloria Inés Ramírez Ríos del PDA ($95.2\%$), y Gustavo Francisco Petro Urrego del PDA ($94.8\%$). Los senadores con mayor probabilidad de tener un punto ideal mayor que $1$ son: María Isabel Mejía Marulanda del PU ($76.9\%$), Jorge Enrique Gómez Montealegre de CV ($71.3\%$), Manuel Guillermo Mora Jaramillo del PU ($70.9.8\%$), Martha Lucía Ramírez de Rincón del PU ($68.7\%$), y Dilian Francisca Toro Torres del PU ($62.2\%$). El senador Jorge Enrique Gómez Montealegre ingresó al Senado de la República a través de CV para reemplazar al senador Jorge de Jesús Castro Pacheco, pero luego de la reforma del 2009 se cambió al PU y en las elecciones del 2010 fue elegido como senador por el PU.

Entre los puntos ideales de los senadores de la coalición resalta la ubicación del punto ideal del senador Carlos Enrique Alvarado Chávez de CR. Se observa que el punto ideal de este senador está ubicado en la zona del espectro político donde se encuentran una buena parte de los puntos ideales del PL. Adicionalmente, la probabilidad de tener un punto ideal menor a $-1$ para el senador Alvarado es de $47.3\%$, la cual es muy superior a $1.59\%$ que es la segunda probabilidad más alta de tener un punto ideal menor a $-1$ entre los senadores de la coalición. Lástimosamente, no hay disponible información relevante sobre la trayectoría política del senador Alvarado que permita entender su comportamiento legislativo durante este periodo. Lo único que podría ayudar a comprender el comportamiento de este senador es la composición y el origen político de CR, el cual albergaba a varios políticos de origen liberal.

Para el PL las probabilidades de tener un punto ideal inferior a $-1$ son a lo sumo del $68\%$. De los 22 senadores del PL, 20 tienen probabilidad inferior al $35\%$ de estar al extremo izquierdo ($91\%$), lo cual es muy diferente para el PDA, donde 9 de los 12 senadores tienen  probabilidad superior al $80\%$ de tener su punto ideal menor a $-1$ ($75\%$). 
El PU y el PC son los partidos de la coalición donde son más altas las probabilidades de tener un punto ideal superior a $1$. En CR es el partido de la coalición donde las probabilidades de estar al extremo derecho del espacio político son más bajas, 15 de los 23 senadores tienen probabilidad inferior al $3\%$ de tener punto ideal superior a $1$ ($65.2\%$). 
Entre los partidos independientes y minorías destaca la probabilidad de tener un punto ideal menor a $-1$ del senador Jesús Enrique Piñacué Achicué de la ASI, la cual es igual a $84.9\%$. Para el resto de senadores de estos partidos las probabilidades de estar en alguno de los extremos son muy bajas.

Algunos miembros de los partidos de la coalición son los que registran mayor probabilidad de estar en el centro del espacio político. En particular, de los 16 senadores que tienen probabilidad mayor al $40\%$ de tener su punto ideal entre $-0.2$ y $0.2$, 9 hacen parte de CR ($56.3\%$) y 3 de AEC ($18.8\%$). Los senadores con mayor probabilidad de tener punto ideal entre $-0.2$ y $0.2$ son: Juan Carlos Restrepo Escobar de CR ($85.7\%$), Antonio del Cristo Guerra de la Espriella de CR ($84\%$), Juan Carlos Martínez Sinisterra de CC ($66.8\%$), Jorge Enrique Vélez García ($64.68\%$), y Rodrigo Lara Restrepo de CR ($63.28\%$). A diferencia de CR y AEC, en CD, el PU, y el PC, las probabilidades de tener punto ideal en el centro para la mayoría de sus senadores es inferior al $20\%$. Por otro lado, para los senadores de los partidos de minorías e independientes, las probabilidades de estar en el centro son casi cero.

\begin{figure}[!t]
    \centering
    \includegraphics[width=14 cm,height=10 cm]{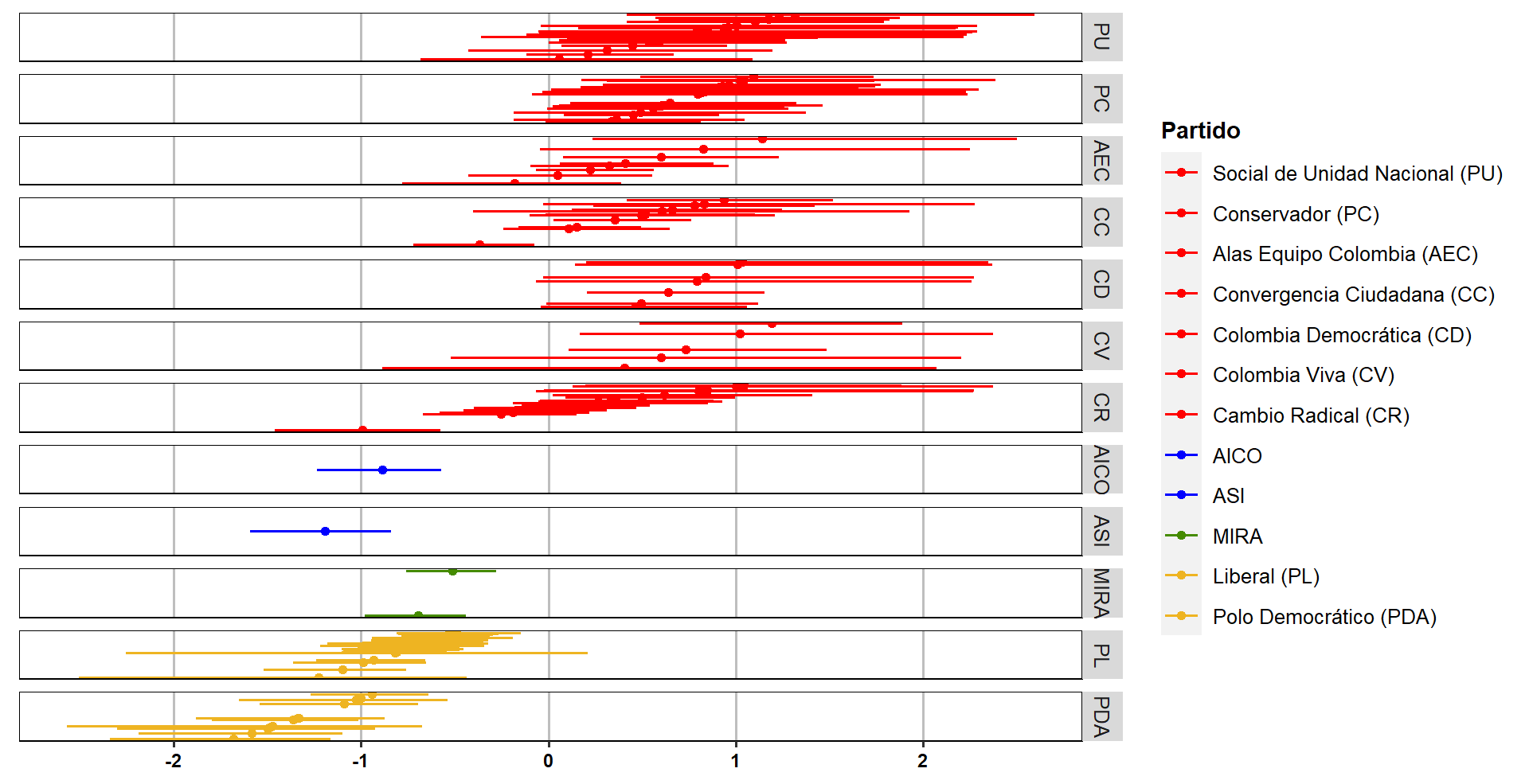}
    \caption{Clasificación por partidos de los puntos ideales estimados de los senadores. Los puntos representan la estimación puntual y las líneas horizontales representan el intervalo de credibilidad simétrico basado en percentiles al $95\%$.}
    \label{credibilidadpartido}
\end{figure}

En la Figura \ref{credibilidadpartido} se muestra el comportamiento de los puntos ideales según el partido político. A grandes razgos se visualizan dos grupos, uno a la izquierda y otro a la derecha del cero. El grupo a la izquierda del cero está conformado por la oposición, los independientes y las minorías, mientras que el grupo a la derecha del cero está conformado por los partidos de la coalición. Dentro de la coalición se observan dos grupos, el primero conformado por los partidos CC, CD, CV, PC y PU, los cuales se encuentran más a la derecha del espacio político, y el segundo conformado por los partidos AEC y CR, los cuales tienen una buena parte de sus puntos ideales ubicados alrededor del centro (cero). La Figura \ref{credibilidadpartido} también muestra la cercanía de algunos puntos ideales de los partidos AEC y CR con algunos puntos ideales del PL, lo cual no es extraño debido al origen político de los partidos AEC y CR. Respecto a la oposición se observa que el PDA se ubica un poco más a la izquierda que el PL, mientras la mayoría del PDA se ubica a la izquierda de $-1$, la mayoría del PL se ubica a la derecha de $-1$. A su vez, las minorías se ubican cerca al $-1$. En cuanto a los independientes, se visualiza que tienen una ubicación en el espacio similar a la del PL.

Lo que se evidencia en la Figura \ref{credibilidadpartido} sugiere que el rasgo latente implícito a la votación nominal del Senado de la República 2006-2010 no es ideológico (izquierda-derecha) debido a que al lado izquierdo se encuentra el PL, que aunque siendo históricamente opuesto al PC, no se ha considerado en los últimos 30 años como un partido de izquierda \citep{uribe2020partidos}. Además, la presencia del MIRA al lado izquierdo, muy cercano al PL reafirma lo enunciado anteriormente, ya que el MIRA ha sido un partido de tendencia centro-derecha. Tampoco parece correcto establecer un rasgo latente no ideológico del tipo oposición-gobierno puesto que los independientes y las minorías no se declararon como parte de alguno de los dos grupos. Por lo tanto la evidencia empírica en este caso sugiere un rasgo latente del tipo oposición-no oposición, donde el espectro político no está dividido de forma equitativa debido al desbalance que hay entre los grupos oposición, minorías, independientes, y coalición. Este mismo patrón se ha observado en Colombia y en otros países de la región cuando se analiza conducta electoral legislativa \citep{londono2014efectos,luque2022bayesian}.

\subsection{Relación entre el escándalo de la parapolítica y el comportamiento legislativo de los senadores}

\begin{figure}[!ht]
    \centering
    \includegraphics[width=12.5 cm,height=18 cm]{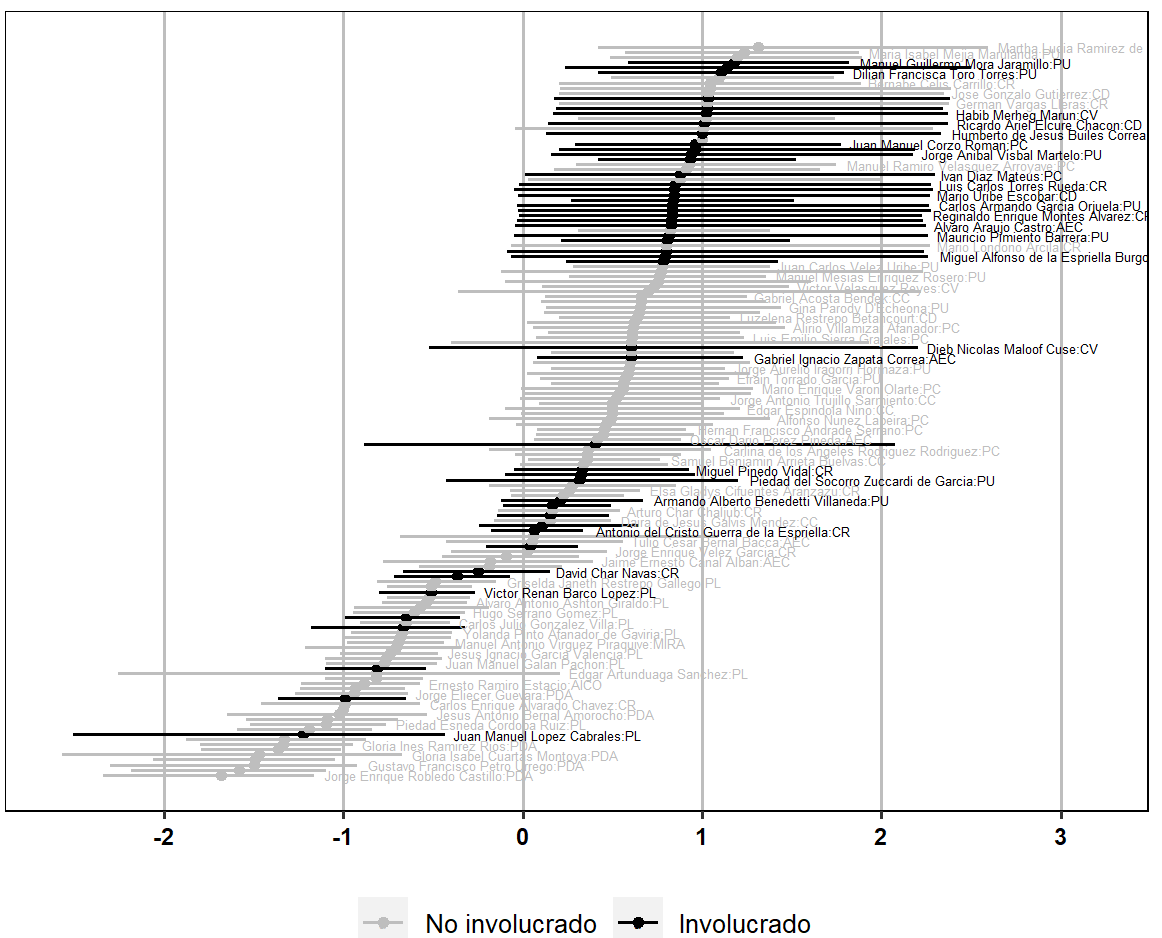}
    \caption{Clasificación de los puntos ideales a partir de si se ha estado o no involucrado en el escándalo de la parapolítica.}
    \label{credibilidadParapolitica}
\end{figure}

En la Figura \ref{credibilidadParapolitica} se contempla los puntos ideales clasificados en dos grupos: involucrados y no involucrados en el escándalo de la parapolítica. La mayoría ($83\%$) de los puntos ideales de los involucrados se encuentran a la derecha del cero. Además, un poco más de la mitad ($57.4\%$) de estos puntos ideales se ubican entre 0.78 y 1.17. La amplitud de los intervalos de credibilidad para una buena parte de los puntos ideales de los involucrados es muy alta, esto debido a la baja participación que tuvieron estos senadores en las votaciones.

Con el fin de establecer si hay relación entre el comportamiento legislativo de los senadores y el estar o no estar involucrado con el escándalo de la parapolítica se lleva a cabo un modelo de regresión logística (frecuentista) donde la variable respuesta ($y$) es si está (éxito) o no está involucrado (fracaso) con el escándalo de la parapolítica y la variable explicativa es el punto ideal ($x$).

El modelo de regresión lógistica implementado es:
$$\textsf{logit}(y)=\beta_0+\beta_1x+\epsilon_i\,,$$
donde $\textsf{logit}(y)=\ln{(y/(1-y))}$ es la función logit, y además, $\textsf{E}(\epsilon_i)=0$ y $\textsf{Var}(\epsilon_i)=\sigma^2$, y $\textsf{Cov}(\epsilon_i,\epsilon_j)=0$, para todo $i$ y todo $j$, con $i\neq j$.

Al realizar el modelo se obtiene que el punto ideal es significativo ($p$-valor = 0.00408) para predecir el estar o no involucrado y a partir de la prueba Chi-cuadrado ($p$-valor = 0.001939) se concluye que el modelo no presenta problemas de ajuste (ver anexo para más detalles). Los estimaciones de los parámetros son:
$$\hat{\beta_0}=-0.9731 \qquad \text{y} \qquad \hat{\beta_1} = 0.7857.$$

La interpretación del coeficiente de regresión del punto ideal indica que un aumento de 1 en el punto ideal de un senador produce que la probabilidad de estar involucrado en el escándalo de la parapolítica sea $e^{\beta_1}=2.19$ veces más alta. Esto quiere decir que entre más a la derecha del espectro político se ubique un senador, más alta será la probabilidad de estar involucrado en el escándalo de la parapolítica.

\begin{figure}[!ht]
    \centering
    \includegraphics[scale=0.63]{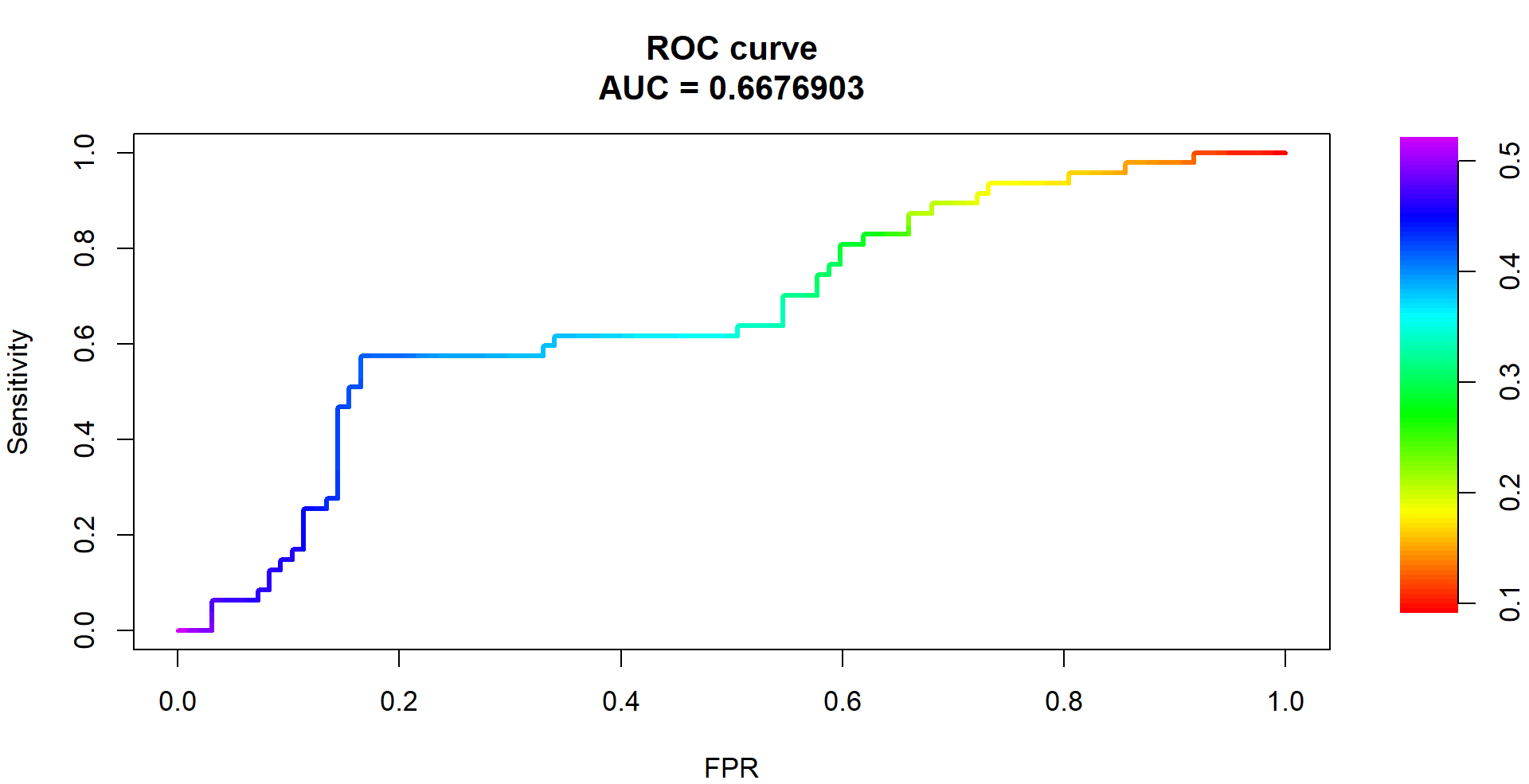}
    \caption{Curva ROC y valor AUC del modelo logístico.}
    \label{curvaRoc}
\end{figure}

Otra forma de evaluar un modelo logístico es mediante el uso del AUC y de la curva ROC. En la Figura \ref{curvaRoc} se muestra la curva ROC y el valor AUC del modelo logístico. La forma de la curva ROC y el valor AUC permiten concluir que el modelo lleva a cabo la clasificación de una mejor forma de la que lo haría un clasificador aleatorio. Se observa que el poder predictivo del clasificador es considerable, sobre todo teniendo en cuenta que se está tomando una sola variable explicativa.

Además, con propósitos ilustrativos, se implementa el modelo logístico Bayesiano usando la distribución normal, con media $\mu=0$ y desviación estándar $\sigma=10$, como distribución previa para los parámetros $\beta_0$ y $\beta_1$. La distribución previa establecida es una previa no informativa. Las estimaciones de los parámetros bajo este modelo son:
$$\hat{\beta_0}=-1.0000 \qquad \text{y} \qquad \hat{\beta_1}=0.8000.$$
Las estimaciones de los parámetros del modelo logístico Bayesiano son muy parecidas a las del modelo logístico frecuentista.

Por todo lo anterior, es posible concluir que existe suficiente evidencia empírica para asegurar que existe 
una relación significativa entre estar o haber estado
involucrado en el escándalo de la parapolítica y el comportamiento legislativo de los senadores del periodo 2006-2010. Esto en parte podría deberse a los objetivos que los paramilitares esperaban conseguir mediante el apoyo que brindaron a varios políticos para ser elegidos en cargos de elección popular. Varios de estos objetivos fueron consignados en un documento que se produjo en una reunión llevada a cabo el 23 de Julio de 2001 en la finca de un jefe paramilitar ubicada en Santa Fe de Ralito, municipio de Tierra Alta, departamento de Cordoba. Este documento se conoce como \textit{el pacto de Ralito} \citep{londono2014efectos}. El pacto de Ralito fue firmado por varios de los jefes paramilitares más importantes de la época y varios políticos, entre los cuales habían alcaldes, gobernadores, concejales, congresistas, futuros aspirantes a cargos de elección popular, etc. Entre los firmantes del pacto de Ralito destacan: Miguel Alfonso de la Espriella, Zulema Jattin Corrales, Juan Manuel López Cabrales, Julio Manzur Abdala, Reginaldo Montes Álvarez, y Piedad Zuccardi, quienes fueron miembros del Senado 2006-2010. Es importante resaltar que varios de los senadores de ese periodo que inicialmente fueron involucrados en el escándalo de la parapolítica, fueron absueltos de sus presuntos vinculos con los paramilitares así que estas personas no estaban bajo la influencia de los paramilitares.

\subsection{Efecto en el cambio de los senadores de anclaje}

Para evaluar la robustez del modelo frente a los senadores de anclaje, se implementa nuevamente el modelo de votación espacial con la elección de dos senadores de anclaje diferentes. Para este caso se fijan las posiciones del senador Gustavo Francisco Petro Urrego del PDA y del senador Mario Uribe Escobar de CD en $-1$ y $1$, respectivamente.

\begin{figure}[!h]
    \centering
    \includegraphics[scale=0.63]{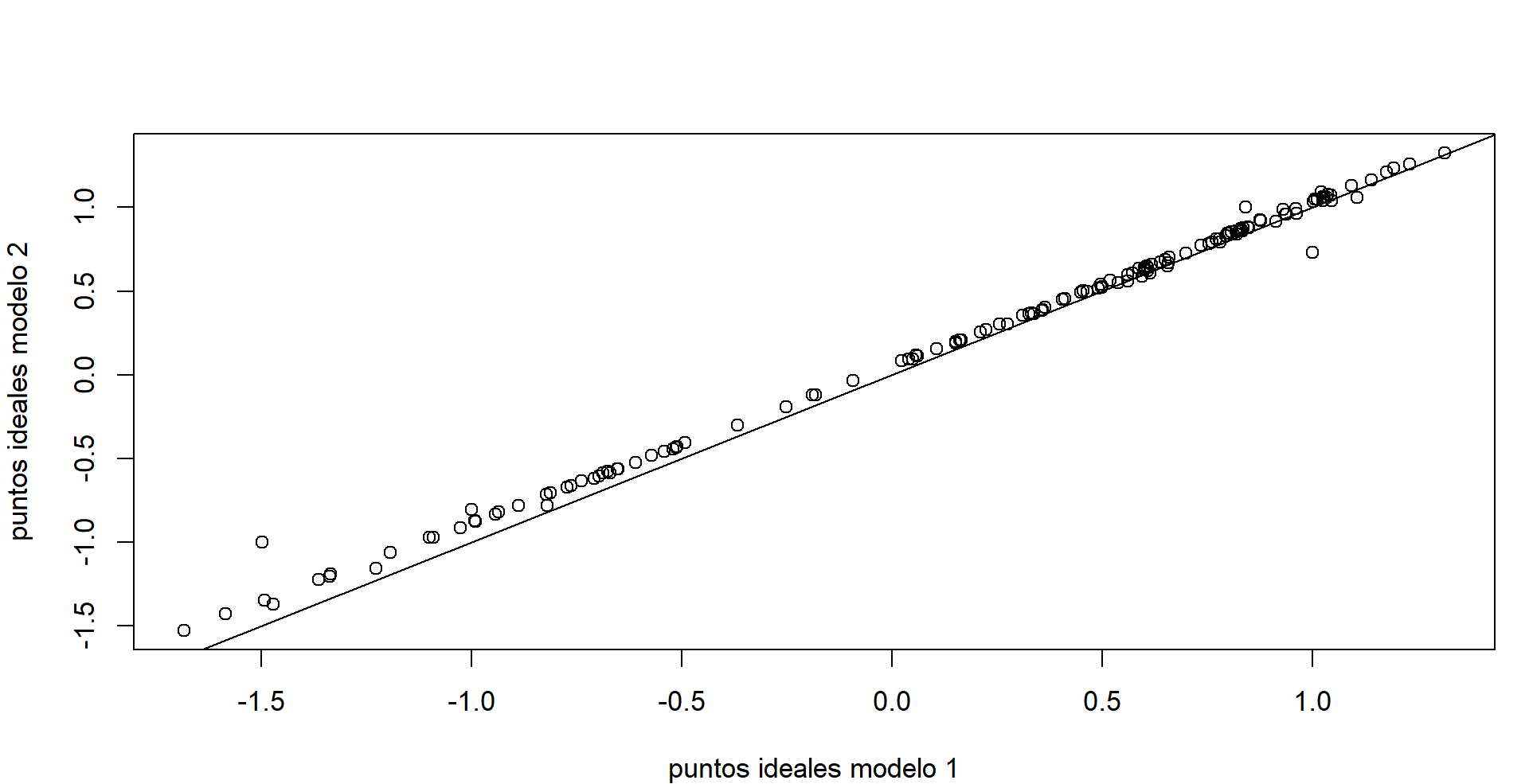}
    \caption{Dispersograma de los puntos ideales estimados de los modelos 1 y 2.}
    \label{dispersograma}
\end{figure}

En la Figura \ref{dispersograma} se muestra el dispersograma de los puntos ideales estimados del primer modelo (modelo 1) contra los puntos ideales del modelo implementado con el cambio de los senadores de anclaje (modelo 2). Se observa que la mayoría de los puntos quedan ubicados cerca de la recta $y=x$. Los puntos ideales negativos son los que varían más de un modelo a otro, mientras que la mayoría de los puntos ideales positivos son muy similares en ambos modelos. La correlación de Spearman de los puntos ideales de los dos modelos es $0.99$.

\begin{figure}[!ht]
    \centering
    \includegraphics[scale=0.63]{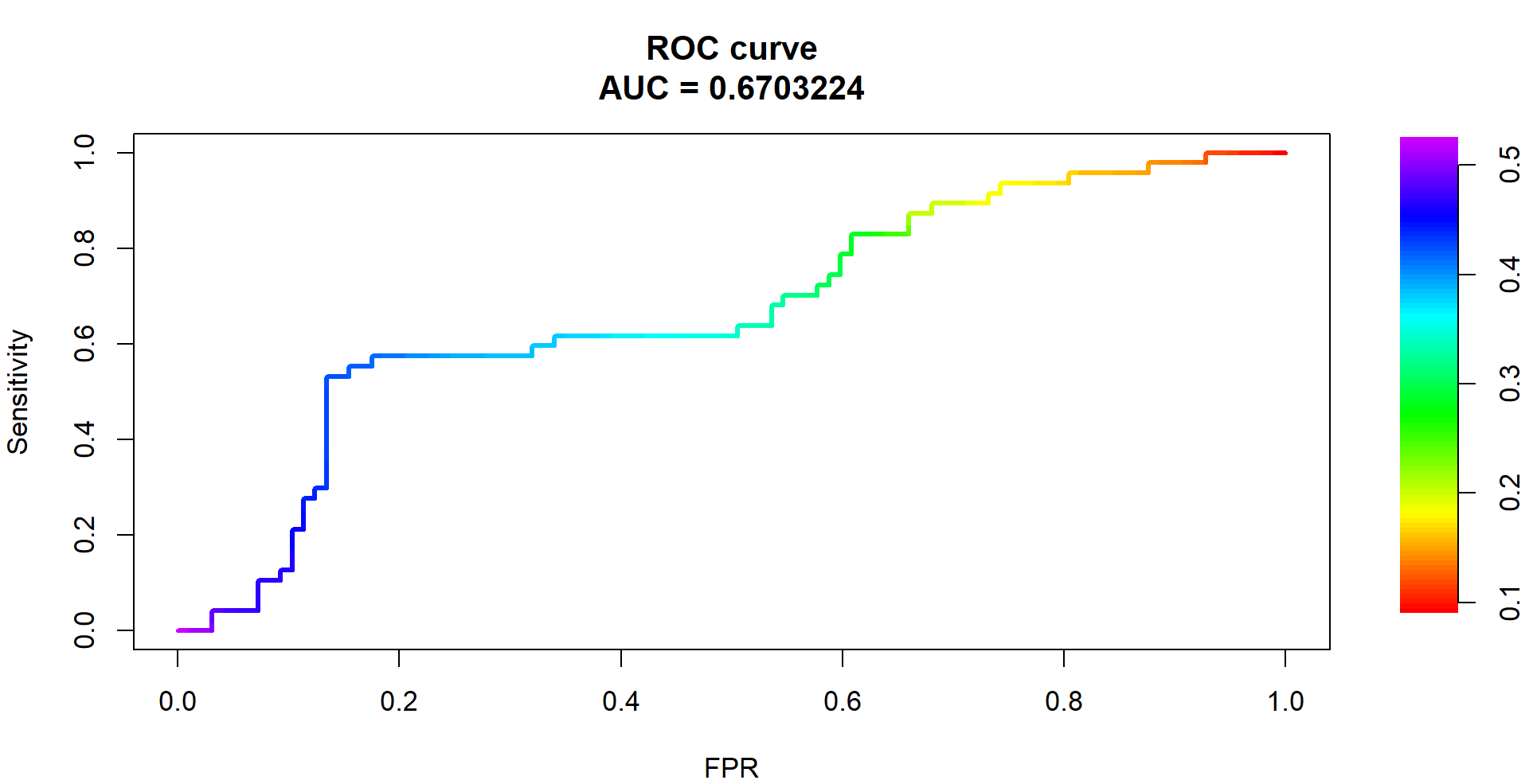}
    \caption{Curva ROC y valor AUC del modelo logístico con los puntos ideales del modelo 2.}
    \label{curvaRoc2}
\end{figure}

Se realiza el modelo de regresión lógistico (frecuentista) con los puntos ideales del segundo modelo. Los parámetros estimados son:
$$\hat{\beta_0}=-1.0111 \qquad \text{y} \qquad \hat{\beta_1}=0.8330.$$
Se observa que los parámetros estimados para ambos modelos son muy similares. Además la prueba Chi-cuadrado arroja un $p$-valor igual a $0.001769$, el cual es muy parecido al $p$-valor del modelo 1. Por último, en la Figura \ref{curvaRoc2} se visualiza la curva ROC y el valor AUC para el modelo logístico con los puntos ideales del segundo modelo. La curva ROC y el valor AUC de los dos modelos logísticos son muy similares. De lo anterior, se concluye que el cambio de los senadores de anclaje no altera los resultados obtenidos acerca de la incidencia del escándalo de la parapolítica en el comportamiento legislativo de los senadores.

\section{Discusión}

Se realizó la consolidación de la base de datos de los senadores y de las votaciones legislativas del periodo 2006-2010. A partir del escándalo de la parapolítica, se clasificó los senadores en dos grupos: involucrados y no involucrados. Esta clasificación permitió realizar un modelo de regresión logística para establecer que hay relación entre estar o haber estado involucrado en el escándalo de la parapolítica y el comportamiento legislativo de los senadores del periodo 2006-2010.

Mediante el cambio de los senadores de anclaje usados para el modelo inicial, se implementó un segundo modelo con el fin de observar posibles cambios abruptos en la estimación de los puntos ideales. Se encontró que la mayoría de las estimaciones de los puntos ideales son muy similares. Además se llevo a cabo una regresión logística con los puntos ideales del segundo modelo y se encontró que los resultados eran muy parecidos a los obtenidos con la primera regresión logística. Lo anterior refuerza la conclusión acerca de la relación entre el escándalo de la parapolítica y el comportamiento legislativo de los senadores.

Se encontró un rasgo latente no ideológico (oposición-no oposición) que subyace a la votación de los senadores, lo cual coincide con el rasgo latente encontrado en \cite{luque2022bayesian}. Adicionalmente, se obtuvo que los grupos de oposición y de coalición son los que mayor variabilidad presentan en su comportamiento legislativo. Esto último es opuesto a lo observado en \cite{luque2022bayesian}.

Algunas de las limitaciones del trabajo fueron: el no poder establecer la comisión a la que pertenecía cada uno de los senadores, la imposibilidad de establecer un análisis con los senadores hallados culpables por vínculos con los paramilitares, debido a que varias de las investigaciones por parapolítica no han concluido, y el no haber encontrado información acerca del senador Carlos Enique Alvarado Chávez de CR, que permitiera entender la ubicación de su punto ideal en el espectro político.

Aunque el escándalo de la parapolítica se destapó hasta 2006, se sabe que los vínculos entre paramilitares y políticos vienen de antes del año 2002 \citep{londono2014efectos}. Con base en lo anterior, a futuro se propone implementar la estimación del punto ideal Bayesiano para la Cámara de Representantes de 2006-2010, la Cámara de Representantes de 2002-2006, y el Senado de 2002-2006. Esto permitiría analizar la relación entre la parapolítica y el comportamiento legislativo de los congresistas de los dos periodos legislativos más polémicos por cuenta de la parapolítica. Adicionalmente, se propone realizar de manera conjunta el ajuste del modelo de puntos ideales y del modelo de regresión logístico. Por último, se propone estudiar los puntos ideales de este congreso con puntos ideales incrustados en espacios no euclidianos motivado por los cambios de partidos políticos \citep{yu2020spherical}.

\bibliography{references}
\bibliographystyle{apalike}

\appendix

\section{Convergencia}

\begin{figure}[!h]
    \centering
    \includegraphics[scale=0.32]{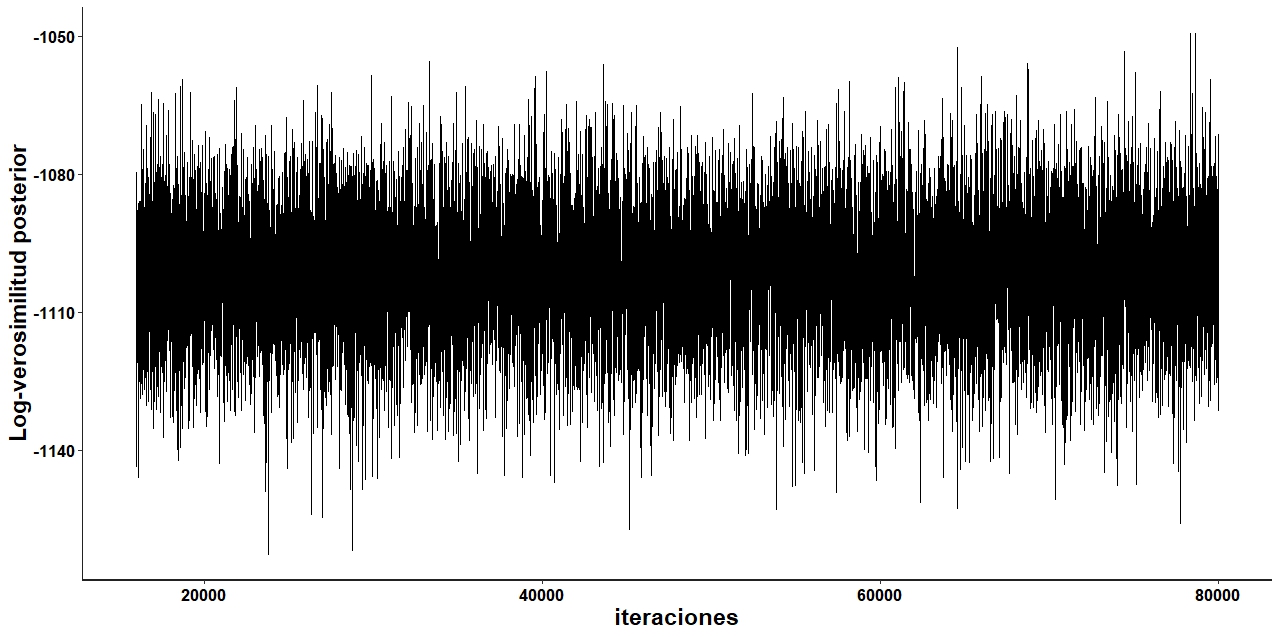}
    \caption{Convergencia del modelo.}
    \label{convergencia}
\end{figure}

La Figura \ref{convergencia} muestra el comportamiento de la log-verosimilitud posterior. Se aprecia que la cadena es estacionaria. 

\begin{figure}[!ht]
 \centering
 \subfigure[]{
 \label{11c}
 \includegraphics[width=0.48\textwidth]{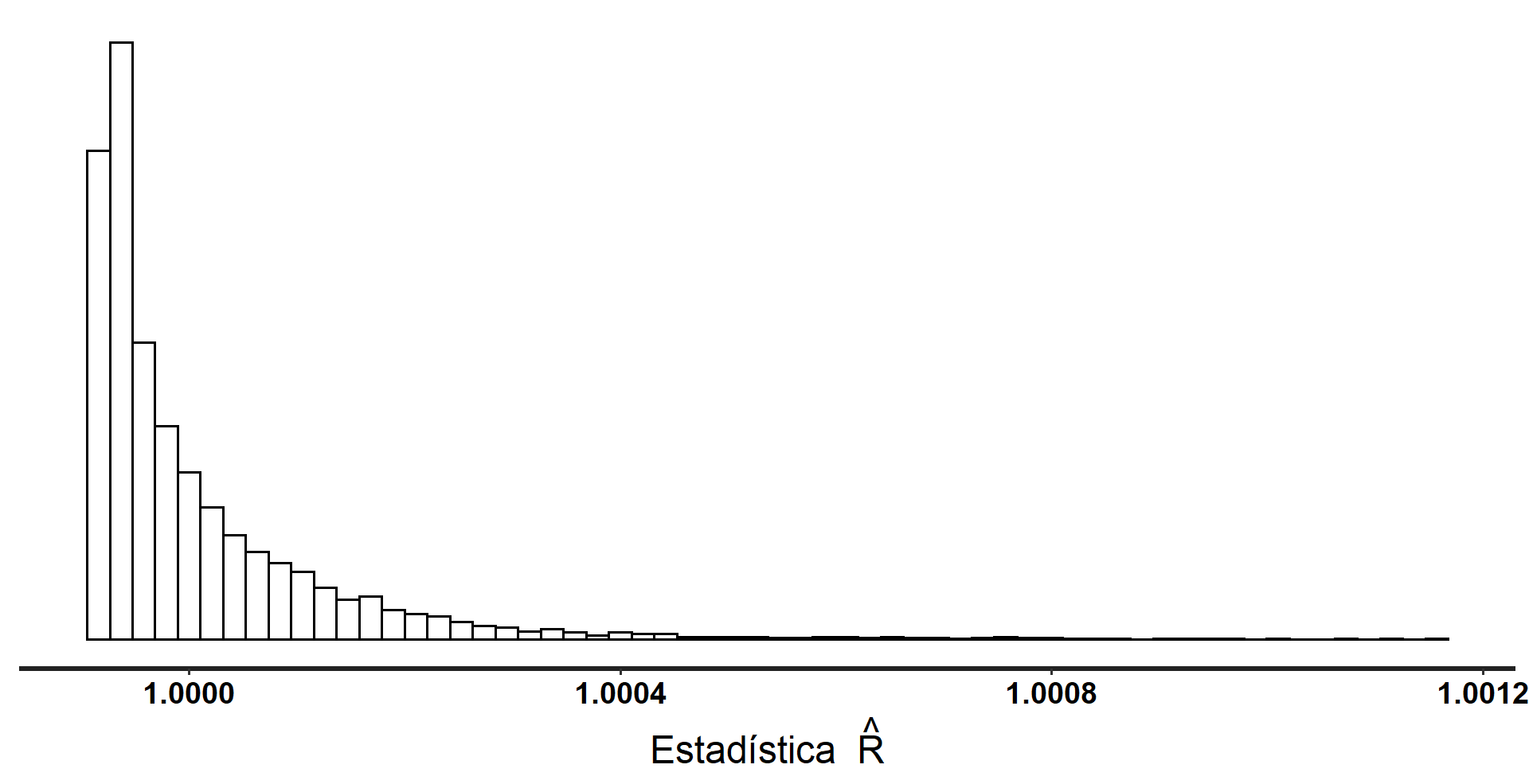}}
 \subfigure[]{
 \label{12c}
 \includegraphics[width=0.48\textwidth]{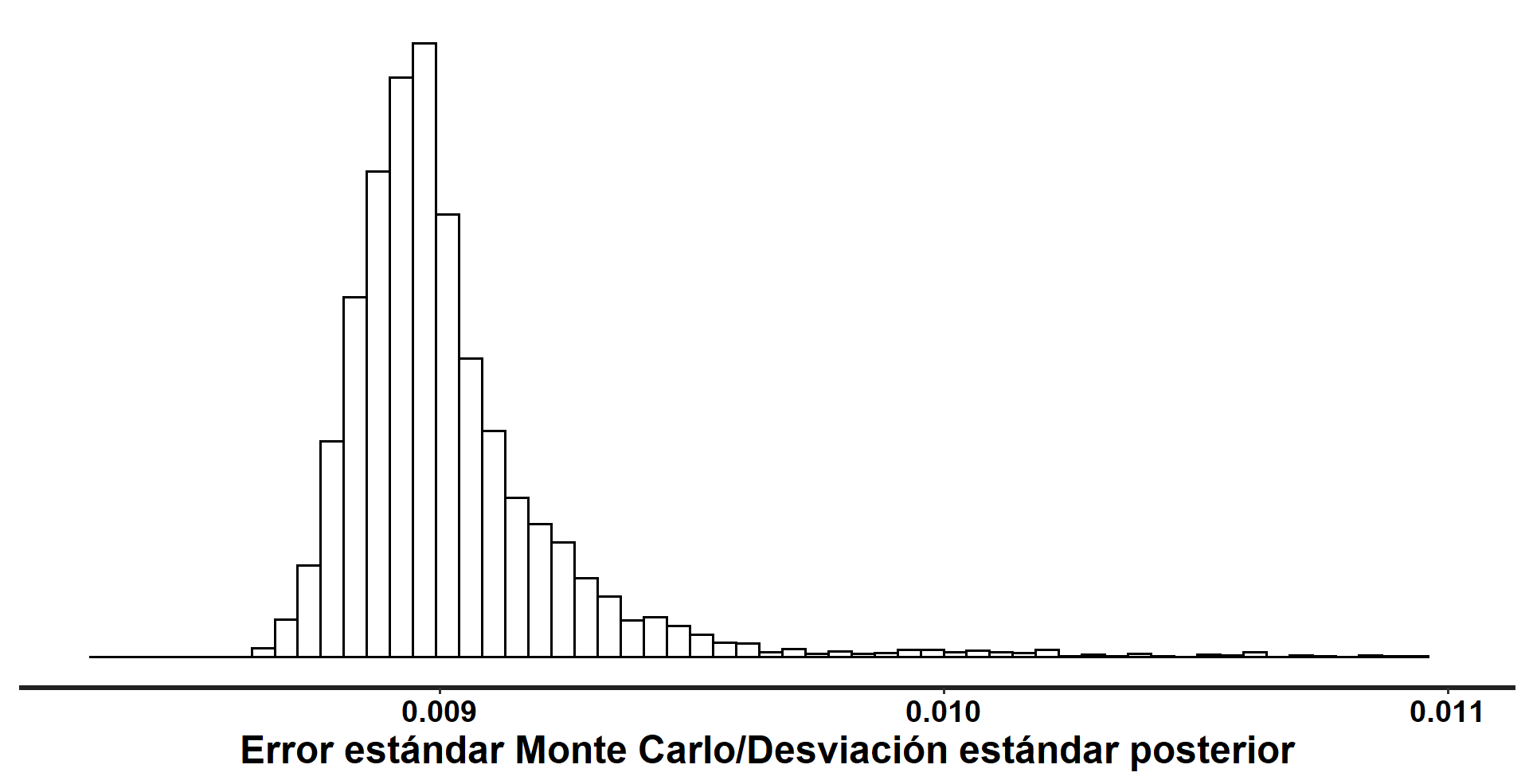}} \\
 \subfigure[]{
 \label{21c}
 \includegraphics[width=0.48\textwidth]{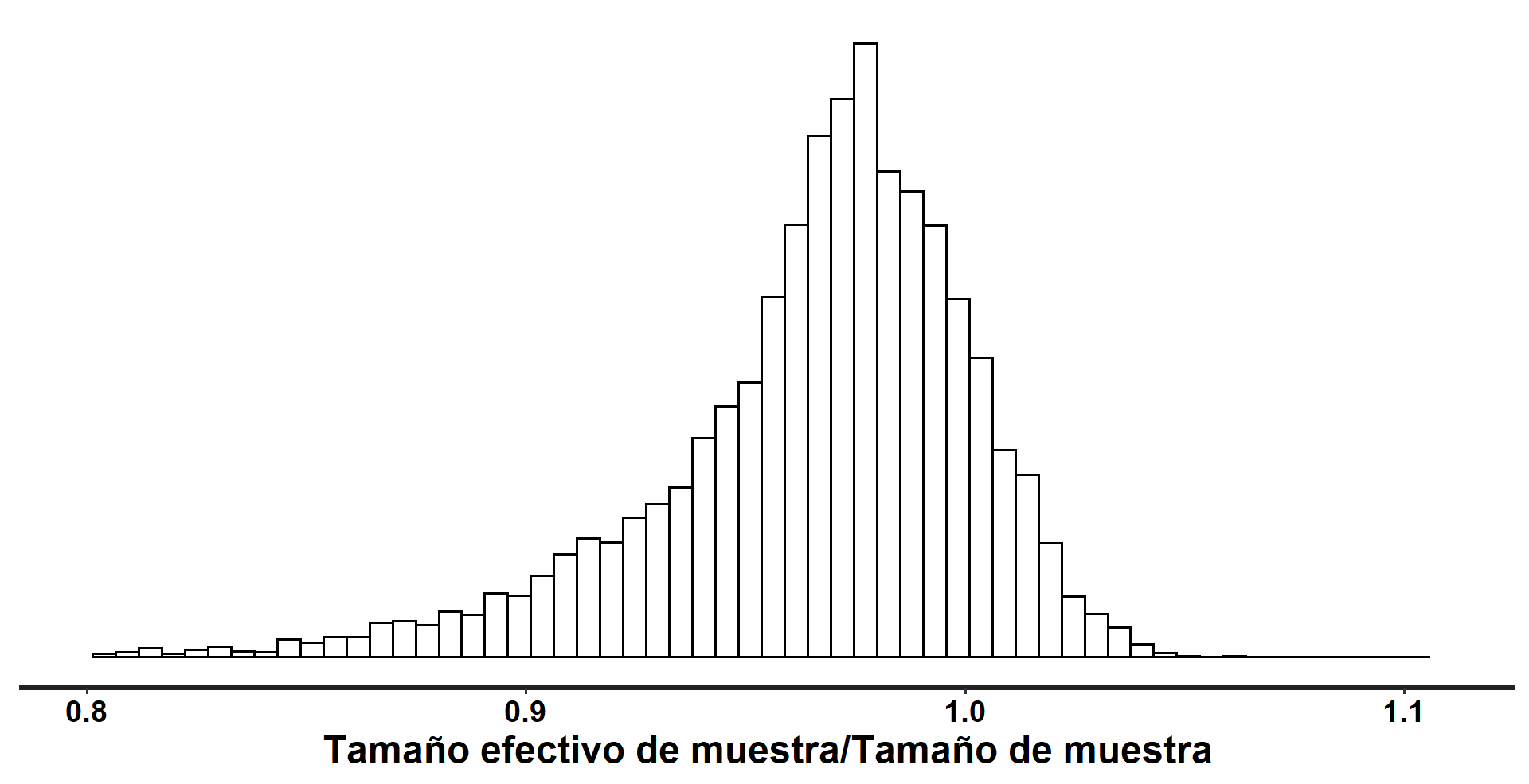}}
 \subfigure[]{
 \label{22c}
 \includegraphics[width=0.48\textwidth]{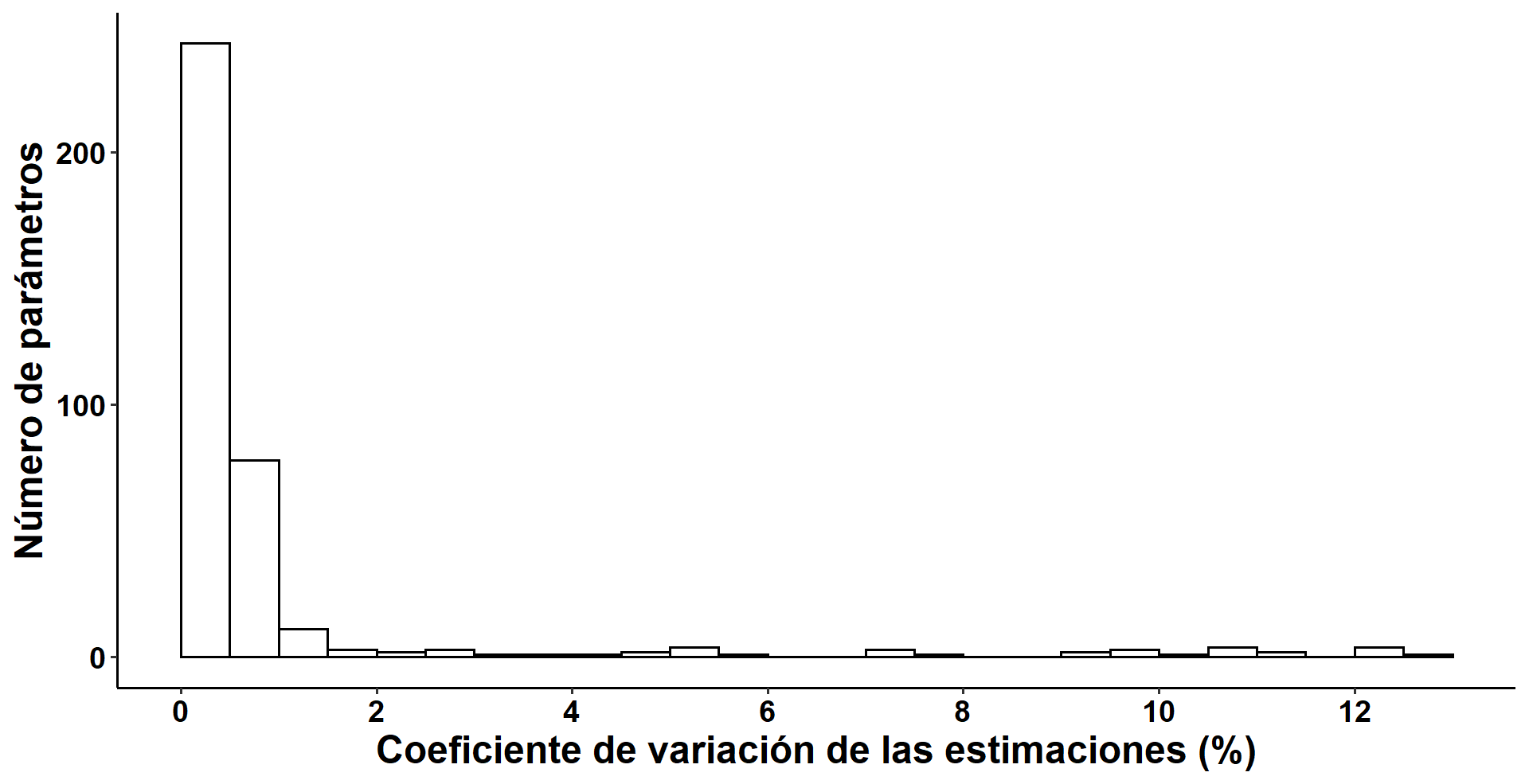}}    
 \caption{Diagnostico de convergencia global para el caso del senado de la República 2006-2010.} 
 \label{convergencia2}              
\end{figure}

En la Figura \ref{11c} se muestra el comportamiento de la estadística $\hat{\textsf{R}}$. Se observa que los valores que toma la estadística $\hat{\textsf{R}}$ son cercanos a $1$, lo cual sugiere que no hay evidencia de falta de convergencia.

En la Figura \ref{12c} se aprecia el cociente entre el error estándar de Monte Carlo y la desviación estándar de las estimaciones. Esta cantidad muestra en su mayoría valores inferiores a 0.01, lo cual sugiere que la fuente de incertidumbre en la estimación de los parámetros
proviene de los datos y no del muestreo de la distribución posterior. 

En la figura \ref{21c} se muestra el cociente entre el tamaño efectivo de muestra y el tamaño de muestra (número de iteraciones) para el caso del Senado. Este cociente exhibe
en su mayoría valores superiores a 0.9, lo cual indica que la cadena de Markov resultante, ofrece una alta
proporción de muestras aproximadamente independientes e idénticamente distribuidas de la posterior, para llevar a cabo el proceso de inferencia sobre los parámetros del modelo.

En la Figura \ref{22c} se muestra el comportamiento del coeficiente de variación para las estimaciones del Senado. El coeficiente de variación oscila entre $0\%$ y $14\%$. La mayoría de los parámetros tienen un coeficiente de variación inferior a $2\%$, lo cual sugiere una alta precisión a posteriori en la estimación de los parámetros del modelo.

Por último, se resalta que los gráficos de la Figura \ref{convergencia2} son muy similares a los obtenidos en \cite{luque2021metodos} para el Senado de la República 2010-2014.

\section{Regresión Logística}

El modelo de regresión logística se realiza en R y se obtiene la siguiente salida:

\begin{verbatim}
Call:
glm(formula = parapolitica ~ PuntoIdeal, family = binomial(link = "logit"), 
    data = listacong)

Deviance Residuals: 
    Min       1Q   Median       3Q      Max  
-1.2110  -0.9731  -0.6379   1.2712   2.0272  

Coefficients:
            Estimate Std. Error z value Pr(>|z|)    
(Intercept)  -0.9542     0.2114  -4.513  6.4e-06 ***
PuntoIdeal    0.7857     0.2736   2.872  0.00408 ** 
---
Signif. codes:  0 ‘***’ 0.001 ‘**’ 0.01 ‘*’ 0.05 ‘.’ 0.1 ‘ ’ 1

(Dispersion parameter for binomial family taken to be 1)

    Null deviance: 181.90  on 143  degrees of freedom
Residual deviance: 172.29  on 142  degrees of freedom
AIC: 176.29

Number of Fisher Scoring iterations: 4
\end{verbatim}

Como el $p$-valor es muy pequeño (igual a 0.00408) se tiene que la variable explicativa punto ideal es significativa.

Se realiza la prueba Chi-cuadrado para ver si el modelo es adecuado:

\begin{verbatim}
Analysis of Deviance Table

Model: binomial, link: logit

Response: parapolitica

Terms added sequentially (first to last)


           Df Deviance Resid. Df Resid. Dev  Pr(>Chi)
NULL                         143     181.90
PuntoIdeal  1   9.6067       142     172.29  0.001939 **
              
---
Signif. codes:  
0 *** 0.001 ** 0.01 * 0.05 . 0.1  1    
\end{verbatim}

El $p$-valor del test likelihood ratio es muy pequeño (igual a 0.001939), luego se puede concluir que el modelo es adecuado.

Se calcula $e^{\hat{\beta}}$ para poder interpretar el coeficiente del punto ideal.

\begin{lstlisting}
> exp(coef(modelo))
(Intercept)  PuntoIdeal 
  0.3851036   2.1938927
\end{lstlisting}

El valor obtenido indica que un aumento de 1 en el punto ideal lleva a que la probabilidad de estar involucrado en parapolítica sea 2.19 veces más alta.

Se evalua el supuesto de linealidad del modelo, el cual consiste en que la relación entre las variables predictoras continuas y la transformación logit de la variable respuesta es lineal. Se usa el test de Box-Tidwell.

\begin{verbatim}
> boxTidwell(logodds ~ listacong$PuntoIdeal)
 MLE of lambda Score Statistic (z) Pr(>|z|)
             1              1.4583   0.1447

iterations =  0
\end{verbatim}

Como el $p$-valor del test de Box-Tidwell no es menor a 0.05, se puede concluir que no se incumple el supuesto de linealidad.

\begin{figure}[!ht]
    \centering
    \includegraphics[scale=0.63]{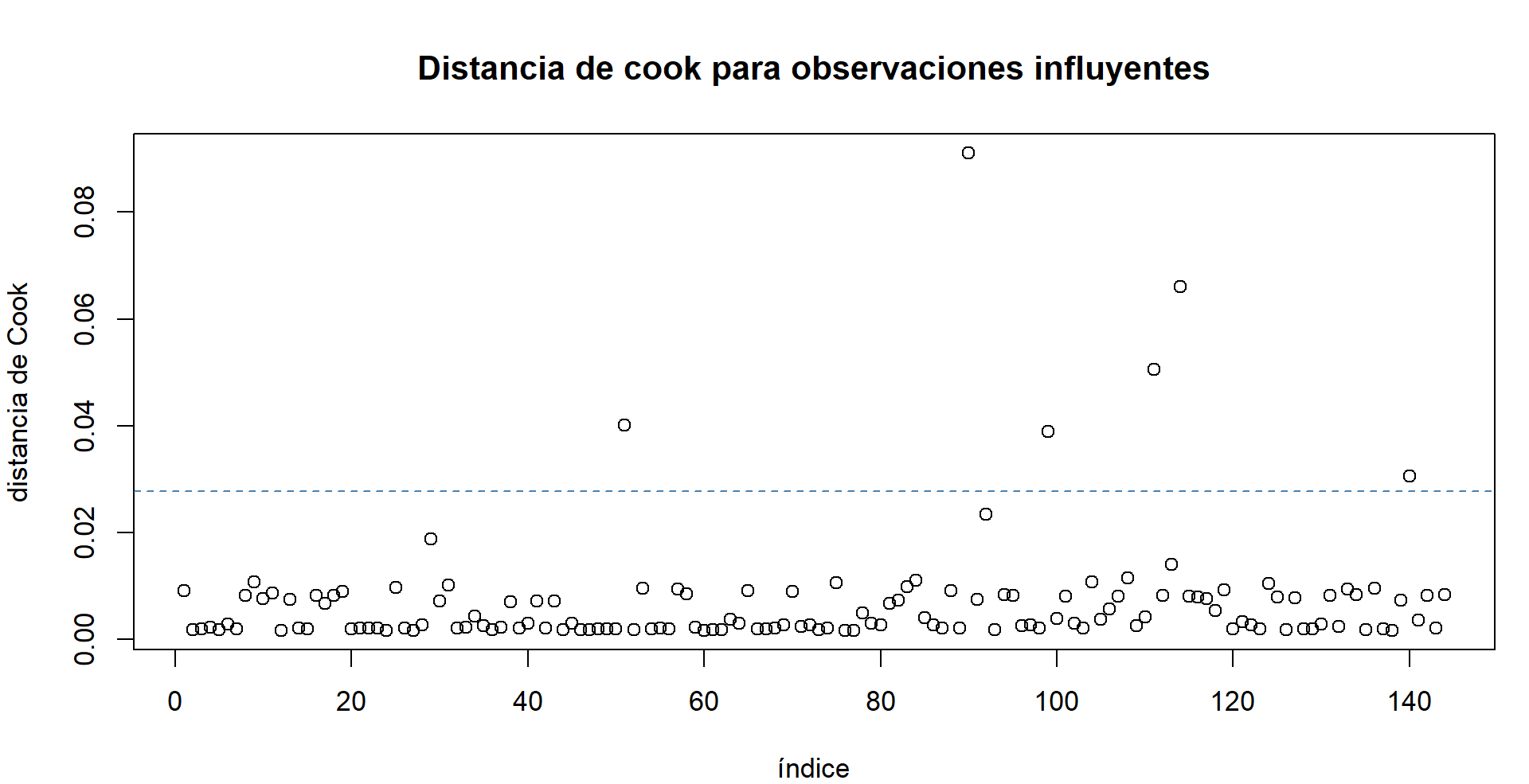}
    \caption{Distancia de cook para identificar observaciones influyentes en el modelo de regresión logística.}
    \label{distanciaCook}
\end{figure}

Por último, se usa la distancia de Cook para identificar puntos influyentes. En la figura \ref{distanciaCook} se aprecia la distancia de Cook para cada uno de los senadores que tienen punto ideal estimado. La línea azul representa el umbral recomendado $4/n$ con el cual se identifican observaciones influyentes ($n$ es el número de individuos, que para este caso son 144). Hay 6 puntos que superan el umbral pero su distancia de Cook no es exageradamente grande. Estos 6 valores corresponden a los puntos ideales de los 6 senadores del PL que estuvieron o han estado involucrados en el escándalo de la parapolítica. Retirar a cada uno de estos puntos ideales del modelo no produjo una mejoría significativa (el AUC sigue siendo similar al del modelo inicial y el coeficiente del modelo no cambia mucho) por lo cual se decide dejar el modelo tal y como estaba.

El modelo de regresión logística Bayesiano se realiza en R y se obtiene la siguiente salida:
\begin{verbatim}
 Model Info:
 function:     stan_glm
 family:       binomial [logit]
 formula:      parapolitica0 ~ PuntoIdeal
 algorithm:    sampling
 sample:       4000 (posterior sample size)
 priors:       see help('prior_summary')
 observations: 144
 predictors:   2

Estimates:
              mean   sd   10%   50%   90%
(Intercept) -1.0    0.2 -1.3  -1.0  -0.7 
PuntoIdeal   0.8    0.3  0.5   0.8   1.2 

Fit Diagnostics:
           mean   sd   10%   50%   90%
mean_PPD 0.3    0.1  0.3   0.3   0.4  

The mean_ppd is the sample average posterior predictive distribution
of the outcome variable (for details see help('summary.stanreg')).

MCMC diagnostics
              mcse Rhat n_eff
(Intercept)   0.0  1.0  2137 
PuntoIdeal    0.0  1.0  2256 
mean_PPD      0.0  1.0  3047 
log-posterior 0.0  1.0  1623 

For each parameter, mcse is Monte Carlo standard error, n_eff is a 
crude measure of effective sample size, and Rhat is the potential 
scale reduction factor on split chains (at convergence Rhat=1).
\end{verbatim}

\end{document}